# MoSSe Janus monolayer as a promising two dimensional material for NO$_2$ and NO gas sensor applications

Rajneesh Chaurasiya and Ambesh Dixit

Department of Physics and Center for Solar Energy, Indian Institute of Technology Jodhpur, 342037, India

[#]**ambesh@iitj.ac.in**

**Abstract:**

Gas sensing mechanism of H$_2$S, NH$_3$, NO$_2$ and NO toxic gases on transition metal dichalcogenides based Janus MoSSe monolayers are investigated using the density functional theory. The pristine and defect included MoSSe layers are considered as a host material for adsorption study. Three types of defects (i) molybdenum vacancy, (ii) selenium vacancy, and (iii) sulfur/selenium vacancy are studied to understand their impact on electronic properties and sensing of these gas molecules. The formation energy is computed to predict the stability of these defects and noticed that selenium vacancy is the most stable among other defects. The adsorption of gas molecules is evaluated in terms of adsorption energy, vertical height, charge difference density, Bader charge analysis, electronic and magnetic properties. The maximum adsorption energy for H$_2$S, NH$_3$, NO$_2$ and NO molecules on pristine Janus MoSSe monolayer are ~ -0.156eV, -0.203eV, -0.252eV, and -0.117eV, respectively. Selenium and sulfur/selenium defects significantly improve the sensing of the gas molecules. NO$_2$ gas molecule dissociates and forms oxygen doped NO adsorption in selenium and sulfur/selenium defect included MoSSe Janus monolayer. The adsorption energy values are ~ -3.360eV and -3.404eV for Se and S/Se defects included MoSSe layer, respectively. Further, the adsorption of NO$_2$ molecule induced about 1μ$_B$ magnetic moment. In contrast, NO molecule showed chemisorption on the surface of the selenium and sulfur/selenium defect included Janus MoSSe monolayers, whereas H$_2$S and NH$_3$ molecules showed physisorption with their adsorption energies in the range of -0.146 to -0.238 eV and -0.140 to -0.281 eV, respectively. The adsorption of H$_2$S, NH$_3$, NO$_2$ and NO molecule on the pristine and defected monolayers suggest that selenium and sulfur/selenium vacancy defects are more prominent for NO$_2$ and NO gas molecule adsorption.

# 1. Introduction

Transition metal dichalcogenides (TMDCs) are most exciting materials in the category of two-dimensional systems because of their wide range of suitable electronic properties such as insulating, semiconducting, semimetallic, metallic and superconducting properties for different applications [1]. The sources of such wide range of electronic properties in TMDCs are the types of crystal symmetry between the transition metal and chalcogen elements and electronic configurations i.e. available electron in the d-orbital of transition metal and their hybridization [2]. Types (direct/indirect) and the bandgap values of TMDCs can be tailored depending on the number of monolayers and also by manipulating strains or doping with foreign elements. Thus, tunable electronic properties will be very useful for nanoelectronic and optoelectronic devices [3]. TMDCs are almost thin, flexible and transparent similar to the graphene with tunable electronic properties. TMDCs based materials, especially $MoS_2$ and its derivatives exhibit excellent electronic properties when scaled down to the sub-nanometer i.e. one monolayer due to the quantum confinement and surface effects in conjunction with indirect (for bulk) to direct (for monolayer) bandgap transition [4]. A monolayer with a large active surface area with more active sites is suitable for efficient energy storage and gas sensing applications [5]. Further, TMDC materials, especially $Mo(S/Se)_2$, and $W(S/Se)_2$ based ultrathin layers are mostly reported for sensing, catalyst and surface-based applications [6]. Sensitivity, selectivity, and interaction of numerous gas molecules ($NH_3$, $NO_2$, NO, $CO_2$, CO, $SO_2$, $H_2S$, $H_2O$, $CH_4$, $N_2$, $O_2$) are explored using both theoretical and experimental investigations [7,8]. These studies suggest that defects, doping elements, strain, and electric field stimuli are also important parameters to enhance the selectivity/sensitivity of a particular gas [9,10,19,11–18]. The sensing performance of gases is also explored by changing the chemical compositions of transition metal and chalcogen element or creating TMDCs monolayer heterostructures.

In recent, Janus MoSSe monolayer is attracting attention, where the half (either top or bottom) of the sulfur atoms are replaced by selenium atom, and thus breaking the out of plan symmetry [20]. Janus MoSSe monolayer is also realized experimentally using chemical vapor deposition [21,22]. Further, the process of sulfurization/selenization in $MoSe_2/MoS_2$ monolayer may lead to vacancy and antisite defects. Further, CVD synthesized Janus MoSSe monolayers are also prone to chalcogen defects. Janus MoSSe monolayer showing potential for nano/optoelectronic applications [20,23,24]. Meng et al. [25] investigated theoretically the effect of the various possible defects in Janus MoSSe monolayer. These vacancy and antisite defects will play a vital role to improve the sensitivity and selectivity of the hazardous gases. The most common point defects in TMDCs based monolayers are chalcogen vacancies, causing the defect states below the conduction band, as a result, limit the electron mobility [26]. However, these defects may be beneficial in enhancing the sensitivity/selectivity of foreign elements in Janus MoSSe monolayers.

Hydrogen sulfide ($H_2S$), ammonia ($NH_3$) and nitrogen dioxide ($NO_2$) are the most common pollutants present in the atmosphere. The emission of $NH_3$ and $NO_2$ gases are continuously increasing from the last few years on the global level [27]. Main sources of these gases are agriculture, fossil fuel, automobile industry, oxidation of atmospheric nitrogen etc. $H_2S$ gas mostly occurs in crude petroleum, natural gas, organic matter and human or chemical sewage [28]. Ammonia gas has a significant impact on the creation of particulate matter and human visibility degradation [29]. Nitrogen dioxide absorbs solar radiation and has an effect on atmospheric visibility and climate change [30]. Thus, such toxic gases ($H_2S$, $NH_3$, and $NO_2$) affect human health and create environmental issues such as acid rain, ozone layer depletion, and the greenhouse effect. Further, these may lead to adverse effects on human health such as breathing difficulties, irritation to nose, eyes, skin, and throat or other body organs.

Considering the importance of detecting such environment pollutants, we considered $H_2S$, $NH_3$, $NO_2$, and NO gas molecules for the present study. $MoS_2$, $MoSe_2$ monolayers showed efficient sensing properties for various toxic gases. Janus monolayer is the derivative of these two systems and thus considered for investigation of adsorption studies for these toxic gas molecules. The different geometric arrangements of gas molecules are explored on all possible active sites in pristine and vacancy defect integrated Janus MoSSe monolayer. Vacancy defects such as molybdenum vacancy ($Mo_V$), selenium vacancy ($Se_V$) and sulfur/selenium vacancy ($S/Se_V$) are considered to understand their impact on sensing of the toxic gas molecules. The sensing of these gas molecules is estimated using adsorption energy, vertical height, charge difference density (CDD), Bader charge analysis and electronic properties. MoSSe Janus monolayer showed the excellent adsorption behavior for NO and $NO_2$ gas molecules. The present work is the first report on the adsorption behavior of toxic gases on MoSSe Janus monolayer as per authors' knowledge and will assist the experimentalists in realizing the gas sensor devices based on MoSSe Janus monolayer and to understand the interaction phenomena of the gas molecule with pristine and defective Janus monolayers.

2. **Computational details:-**

Structural, electronic and adsorption properties of Janus MoSSe monolayer are studied using the Density functional theory (DFT) as implemented in quantum espresso [31,32]. A 4x4 supercell is considered for MoSSe Janus monolayer with 15Å vacuum along the c-axis to avoid the interlayer interaction. Generalized gradient approximation (GGA) predicted by Perdew-Burke-Ernzerhof (PBE) is used as an exchange-correlation functional [33] in the present calculations. Ultrasoft pseudo-potentials are considered with plane wave cutoff energy 50Ryd and sampling of the supercell is done using the Monkhorst-Pack [34] scheme with 6x6x1 *K*-points are used for relaxation and 12x12x1 *K*-points used for the total energy and other physical properties calculations. Energy convergence cutoff $1x10^{-8}$ eV is used for self-consistent field (SCF) calculations. Atomic positions are relaxed by minimizing the force between the atoms using Broyden-Fletcher-Goldfarb-Shenno (BFGS) minimization scheme with $1x10^{-3}$ eV/Å force convergence limit. Van der Waals (vdW) force are also considered by adopting the DFT-D2 method proposed by Grimme et al.[35]. $Mo_V$, $Se_V$, and $S/Se_V$ vacancies are created by removing atoms from the supercell in Janus MoSSe monolayer and optimized by minimizing the force on each atom. Charge transfer between the monolayer and gas molecules are carried out using the Bader charge analysis [36].

## 3. Results and Discussions:-

Janus MoSSe monolayer is a three-layer structure, where top layer consists of selenium atoms, the middle layer is made of molybdenum atoms and bottom layer consists of sulfur atoms, shown schematically in Fig.1 (a). Such Janus structures are experimentally synthesized recently using plasma enhanced chemical vapor deposition method [21,22] and thus, important to understand their physiochemical properties and potential for possible applications. The optimized lattice parameter of Janus MoSSe monolayer is 3.249Å, which is in between the lattice parameters of $MoS_2$ and $MoSe_2$ monolayers [21]. We also optimized the structural parameters like the bond length and bond angle, which are marked in Fig.1 (b). The optimized parameters are in agreement with the previously reported results [37]. Further, the structural stability is supported by computing the phonon band dispersion and is shown in Fig. 1(c). The nine vibrational modes are noticed with three acoustic and six optical modes. The noticed positive frequencies for all these vibrational modes support the thermodynamic stability of Janus MoSSe monolayer, consistent with previous reports [38,39]. We also considered chalcogen (sulfur and selenium) and molybdenum defects in Janus MoSSe monolayer. The structural stability of the defect containing monolayers is defined using the formation energy as $E_f = E_{defect} - E_{pristine} + E_{removed}$; where, $E_{defect}$ and $E_{pristine}$ are the total energies for defect containing monolayer and pristine monolayer respectively, and $E_{removed}$ is the total energy of the removed atom [40]. Formation energy for $Mo_V$, $Se_V$ and $S/Se_V$ are about 6.98 eV, 3.16 eV and 6.23 eV, respectively. Thus, $Se_V$ containing MoSSe Janus monolayer is relatively more stable among the investigated defects, as the required formation energy for $Se_V$ defect is the lowest as compared to other defects. This is in agreement with the common observation that the chalcogen atom mediated vacancy defects are more stable in TMDCs based monolayers [41].

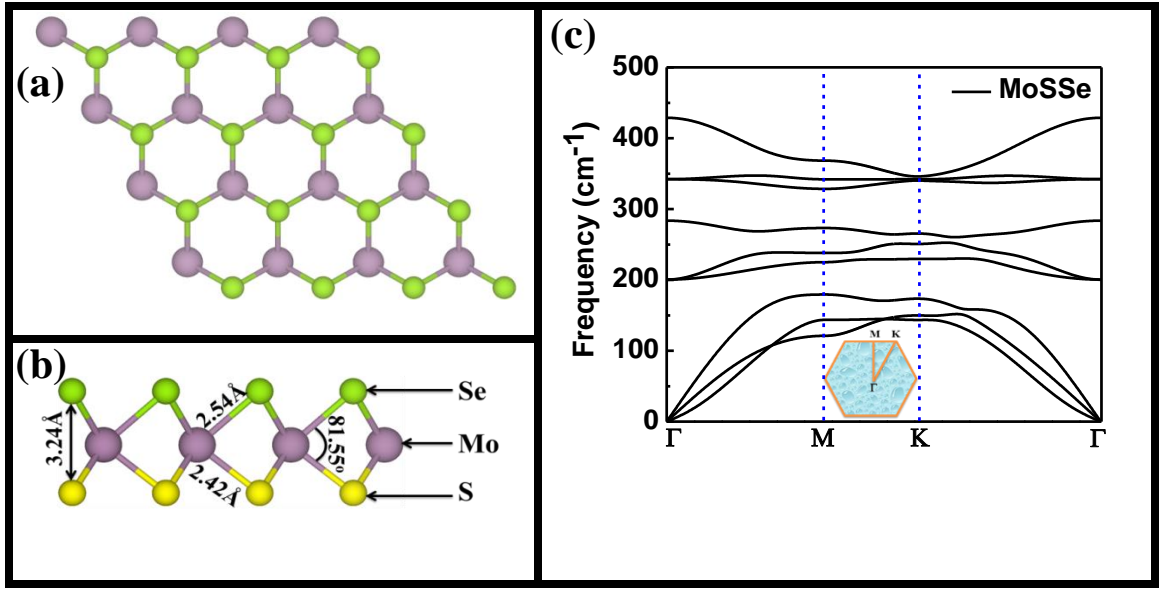

Fig.1 MoSSe monolayer (a) top view [Green, purple and yellow color are representing selenium, molybdenum and sulfur atoms, respectively] (b) side view with optimized structure parameters like the bond length and bond angle, and (c) phonon band structure of pristine Janus MoSSe monolayer with inset showing the hexagonal Brillouin zone in conjunction with high symmetry points.

Further, spin polarized band structure and total density of states (TDOS) for thermodynamically stable MoSSe Janus monolayer are computed and shown in Fig. 2(a) for Janus MoSSe monolayer. Here, valence band minima (VBM) and conduction band maxima (CBM) are located at K-point and symmetry in the spin up and spin down states, confirming the non-magnetic direct bandgap semiconductor behavior with the band gap value ~ 1.58 eV. The noticed band gap value lies in between that of $MoS_2$ and $MoSe_2$ monolayers, consistent with computed values using GGA-PBE exchange-correlation potentials [42]. We also computed partial density of states (PDOS) and shown in Fig. 2(b). VBM and CBM are dominated by Mo-d orbitals, whereas the deeper region of the valence band is mainly dominated by S-p orbitals. S/Se-p orbitals are also contributing to both valance band (VB) and conduction band (CB), hybridizing strongly with Mo-d orbitals as confirmed from the observed charge sharing charge between Mo and S/Se atoms. This gives rise to the formation of covalent bonds between them. The results are consistent with Peng et al. [43]. Further, both experimental and theoretical reports support the observation of chalcogen vacancy defects in MoS$_2$ and MoSe$_2$ monolayers [26,41,44,45]. The similar observations are also noticed and reported based on the formation energy of chalcogen vacancy included TMDCs based monolayer [46–48]. The formation energy of chalcogen vacancy included monolayers is less than the other defects, that's why the observation of chalcogen vacancy defects is more common. Spin polarized band structure in conjunction with TDOS and PDOS for Mo$_V$ defect included MoSSe Janus monolayer is shown in Fig. 2( c & d). Here, the defect states lie near to Fermi energy in VB and CB. These defect states are mainly originating from the unsaturated bond of the S or Se atom, confirmed from the PDOS analysis. TDOS and PDOS show the symmetry in the spin states, confirming the non-magnetic semiconducting behavior. These defects may provide the active sites for adsorption of the toxic gas molecules in MoSSe Janus monolayer.

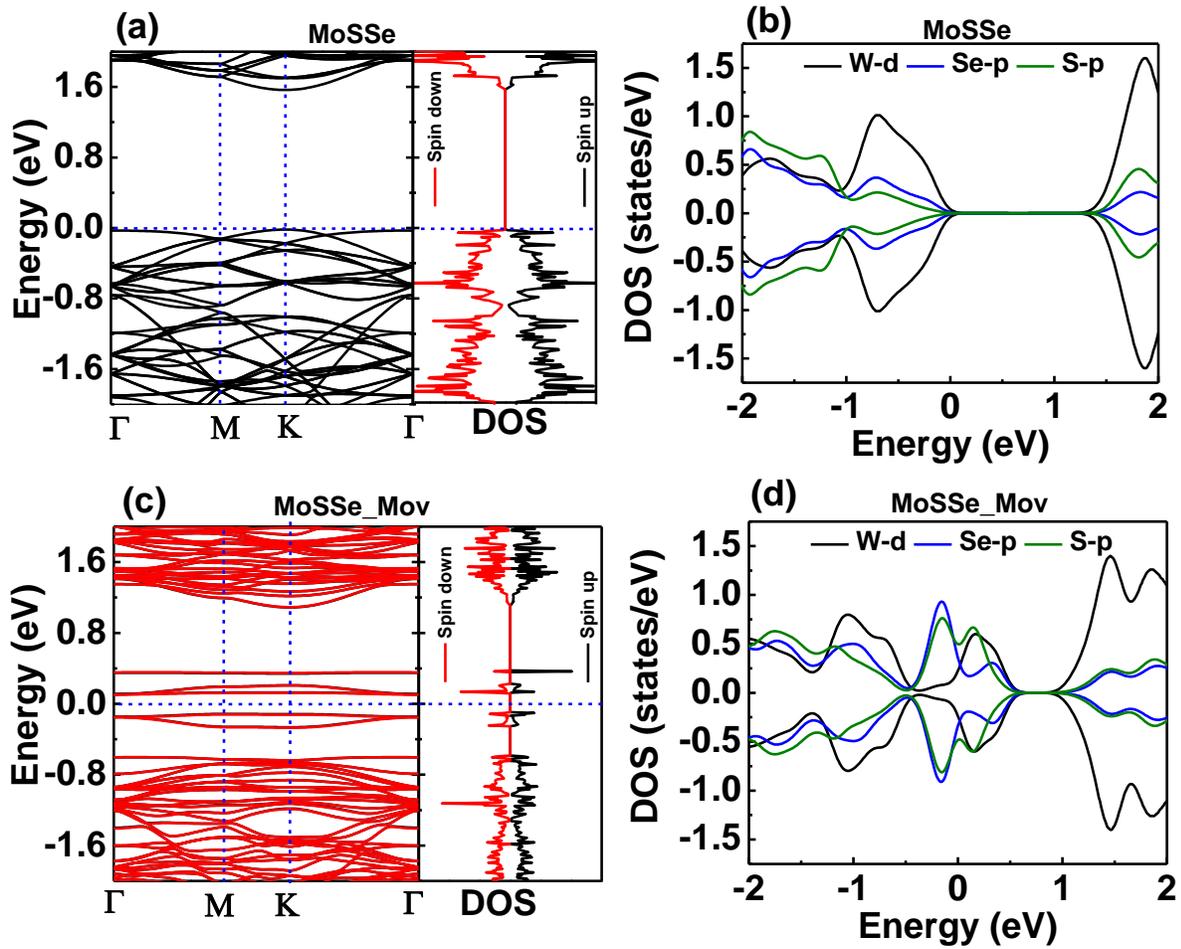

Fig. 2 (a) Spin polarized band structure with TDOS [Black and blue color represent the spin up and spin down, respectively]; (b) PDOS for pristine Janus MoSSe monolayer; (c) spin polarized band structure with TDOS (d) PDOS of Mo$_V$ defect included Janus MoSSe monolayer

The spin polarized band structure with TDOS for Se$_V$ defect included Janus MoSSe monolayer is shown in Fig. 3(a). Here, VBM of Se$_V$ defect included monolayer is observed at Γ point while CBM is located at K point, signifying the indirect band gap (~1.54 eV) semiconductor behavior. The energy difference in VBM at K and Γ point of Brillouin zone is very less and that may be the region of dominating direct bandgap like absorption in such monolayer. The defect states lie below the conduction band around 1 eV due to the unsaturated bonds of the molybdenum atom near the vacancy defect sites. PDOS spectra of Se$_V$ included MoSSe monolayer is shown in Fig. 3(b). VBM and CBM both consist of Mo-d orbitals; however, the defect states mainly originate from the unsaturated molybdenum atomic bonds. The spin polarized band structure and PDOS for Se$_V$ defect included MoSSe monolayer are similar to that of S$_V$ defect included in MoS$_2$ monolayer [46], confirm the non-magnetic behavior. The S$_V$ vacancy defect in Janus MoSSe monolayer gives rise to the n-type semiconducting behavior because the two electrons in Mo atom remain unbonded due to Se$_V$ vacancy and thus contributing to the conduction band electrons.

The spin polarized electronic band structure for S/Se$_V$ defect included Janus MoSSe monolayer is shown in Fig. 3(c). VBM and CBM are located at Γ and K points, respectively, supporting the indirect bandgap semiconductor behavior with a band gap value ~ 1.37eV. S/Se$_V$ defect in MoSSe monolayer gives rise to the defects states ~ 0.8 eV below the CBM, similar to that of S$_{2V}$ defect in MoS$_2$ monolayer, where also the defect states lie below the CBM with reduced band gap with respect to pristine MoS$_2$ monolayer [47]. PDOS spectra of S/Se$_V$ defect included MoSSe monolayer shows that the defect states are due to the d orbital of molybdenum which is similar to that observed for Se$_V$ defect included Janus MoSSe monolayer, except that the defect states are shifted around 0.2eV below as compared to that of Se$_V$ defect included Janus MoSSe monolayer. Thus, S/Se$_V$ defect included monolayer will also exhibit similar n-type behavior. Janus MoSSe monolayer is the derivative of both MoS$_2$ and MoSe$_2$, and thus, provides sulfur and selenium planer sides in a single material. This makes it unique for investigating the adsorption behavior the toxic gases.

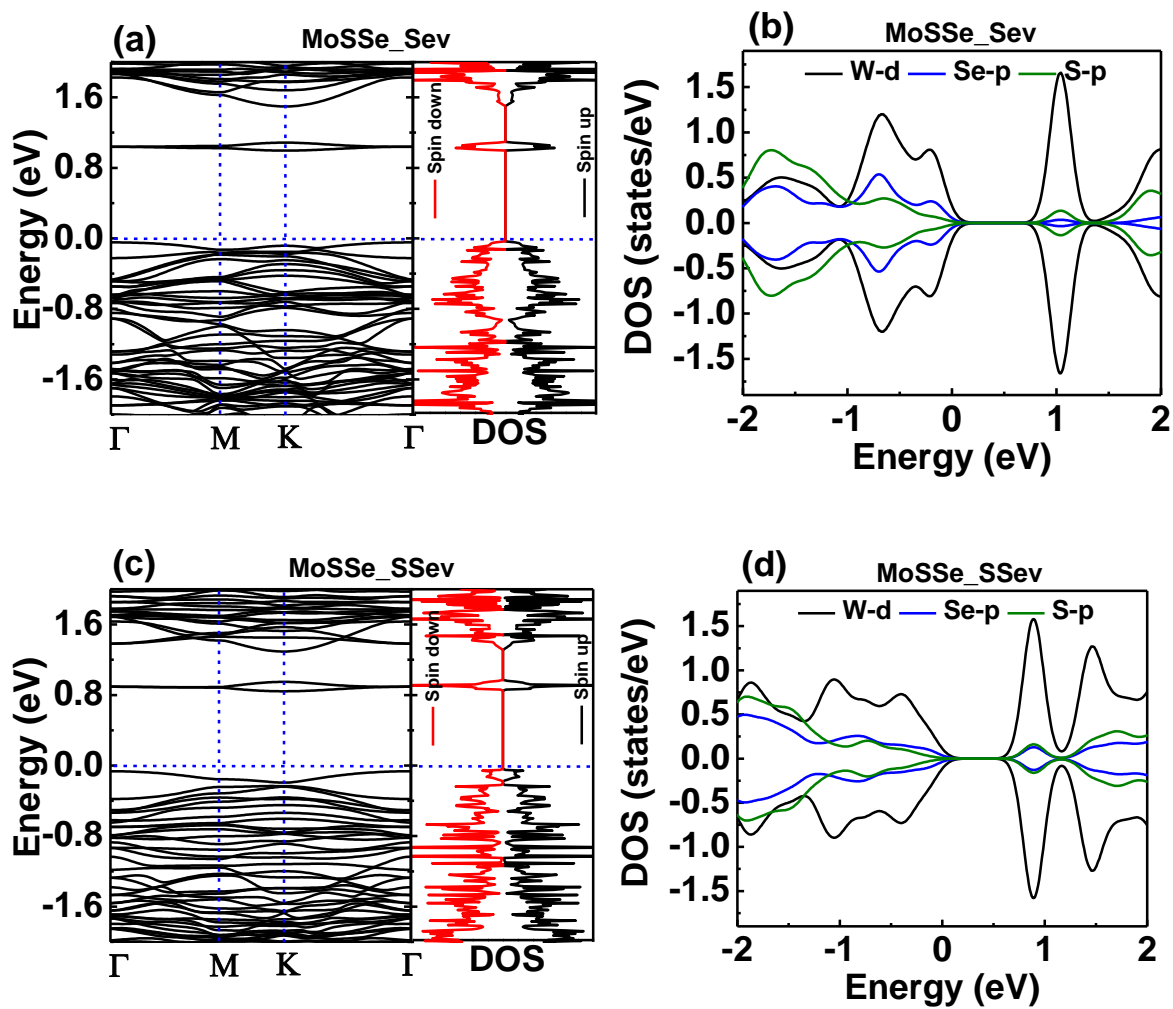

Fig. 3 (a) Spin polarized band structure with TDOS [black and blue color represent the spin up and spin down contributions, respectively]; (b) PDOS of $Se_V$ defect involve in Janus MoSSe monolayer; (c) spin polarized band structure with TDOS (d) PDOS of $SSe_V$ defect involve in Janus MoSSe monolayer

**Adsorption behavior:**

Further, to investigate the sensing properties of pristine and defect included MoSSe Janus monolayers, we considered $H_2S$, $NH_3$, $NO_2$ and NO toxic gas molecules in this study with an objective to find the most suitable binding sites for these molecules. We considered different orientations of these molecules with pristine and defect included MoSSe Janus monolayer to understand the adsorption mechanism. We optimized these structures without any constraint using DFT-D2 correction to include the vdW force for computing the accurate adsorption energies [8]. The stability of adsorbed gas molecules is estimated using the adsorption energy as $E_{ad} = E_{monolayer+mol} - E_{monolayer} - E_{mol}$ where $E_{monolayer+mol}$ and $E_{monolayer}$ are the total energy of molecule adsorbed monolayer and without molecule adsorbed monolayer, respectively and $E_{mol}$ is the total energy of the gas molecule [49]. Considering the geometrical configurations due to two types of atoms in these gas molecules, we investigated adsorption of individual molecules from each atomic side, as shown in Fig. 4. The negative values of adsorption energy signify that the process is energetically favorable [50]. This criterion is considered in evaluating the adsorption behavior of toxic gas molecules.

We analyzed the adsorption of toxic gas molecules by investigating the electronic properties in terms of spin polarized band structure. Further, a gas sensor works on the principle of change in electrical conductivity during the gas adsorption because of the charge transfer between the adsorbent and host material. This is also considered as a parameter in evaluating the sensing characteristics. The electrical conductivity of the semiconductor depends on the band gap of semiconductor material as $\sigma \propto e^{\frac{-E_g}{2KT}}$; where $E_g$, $K$ and T are the band gap, Boltzmann constant and T temperature, respectively [51]. The recovery time $\tau$ of a sensor depends on the adsorption energy as $\tau \propto \exp\left(-\frac{E_{ad}}{KT}\right)$; $E_{ad}$ is the adsorption energy, and $K$ is Boltzmann's constant. In conjunction with adsorption energy, vertical height (the minimum distance between the lower atom of adsorbent and top atom of host material), Bader charge analysis, CDD, electronic and magnetic properties are also considered as the parameters to evaluate the sensing performance. Adsorption energy defines the interaction of the adsorbent with the host. The charge transfer mechanism is a basic working mechanism in two-dimensional materials based gas sensor. The adsorbent may behave like a donor or acceptor and thus, receive carrier from the host or inject carrier to the host, leading to the charge transfer between adsorbate and host [9]. The direction of charge transfer may reflect in the band structure and finally on the electrical conductivity. The CDD of the adsorbed gas molecule are computed using the equation $\Delta \rho = \rho_{monolayer+gas} - (\rho_{monolayer} + \rho_{gas})$, where, $\rho_{monolayer+mol}$, $\rho_{monolayer}$ and $\rho_{mol}$ are the charge density of gas molecule included the monolayer, without gas molecule adsorbed monolayer and gas molecule, respectively.

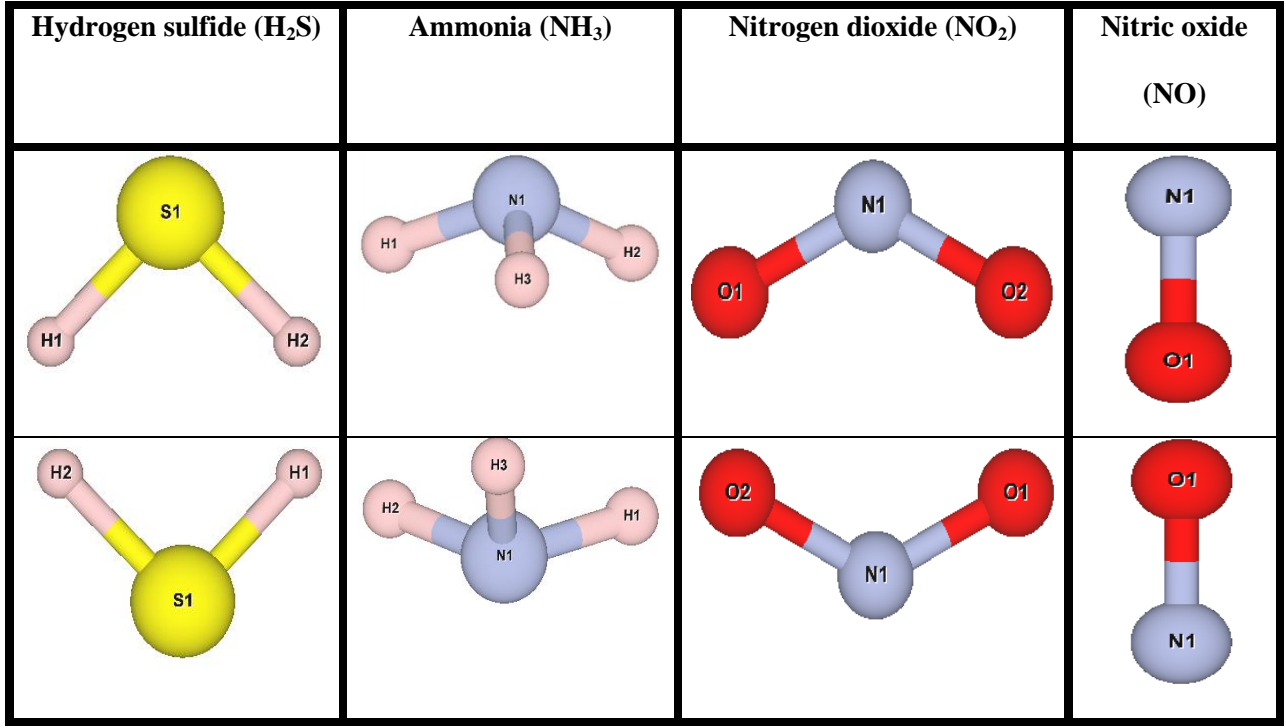

Fig.4 Schematic representation of the initial orientation of H₂S, NH₃, NO₂ and NO gas molecule used for adsorption, in which the top panel and bottom panel represent one configureation 'configuration' and the other configuration 'configuration1', respectively. [pink, yellow, grey, and red represents the hydrogen, sulfur, nitrogen, and oxygen atoms, respectively]

**Table1.** The electronic behavior (SC, D and I represent the semiconductor, direct bandgap, and indirect bandgap, respectively), charge transfer between the monolayer and molecule (Negative means charge transfer from molecule to monolayer and positive means charge transfer from monolayer to molecule), magnetic moment, adsorption energy, vertical height and bond lengths for H₂S, NH₃, NO₂ and NO gas molecules.

| Adsorption configuration of the gas molecule | | Electronic Behavior | Charge (in units of e) | Magnetic moment | | Adsorption energy (eV) | Vertical height (Å) | Bond length of the molecule | |
|---|---|---|---|---|---|---|---|---|---|
| | | | | Before Adsorption | After Adsorption | | | Before Adsorption | After Adsorption |
| H$_2$S | P | SC (I) | +0.011 | 0.00 | 0.00 | -0.156 | 2.292 | 1.353 | 1.355 |
| | P1 | SC (D) | -0.001 | 0.00 | 0.00 | -0.147 | 3.019 | 1.353 | 1.354 |
| | Mov | SC (D) | +0.010 | 0.01 | 0.00 | -0.232 | 2.074 | 1.353 | 1.355 |
| | Mov1 | SC (D) | +0.002 | 0.01 | 0.00 | -0.146 | 3.103 | 1.353 | 1.353 |
| | Sev | SC (I) | +0.017 | 0.00 | 0.00 | -0.232 | 1.572 | 1.353 | 1.355 |
| | Sev1 | SC (I) | -0.003 | 0.00 | 0.00 | -0.222 | 2.044 | 1.353 | 1.355 |
| | SSev | SC (I) | +0.018 | 0.00 | 0.00 | -0.238 | 1.609 | 1.353 | 1.355 |
| | SSev1 | SC (I) | -0.002 | 0.00 | 0.00 | -0.222 | 2.128 | 1.353 | 1.354 |
| NH$_3$ | P | SC (I) | -0.005 | 0.00 | 0.00 | -0.140 | 2.379 | 1.022 | 1.022 |
| | P1 | SC (D) | -0.028 | 0.00 | 0.00 | -0.203 | 2.559 | 1.022 | 1.023 |
| | Mov | SC (D) | -0.010 | 0.01 | 0.01 | -0.146 | 2.499 | 1.022 | 1.022 |
| | Mov1 | SC (D) | -0.004 | 0.01 | 0.01 | -0.160 | 2.484 | 1.022 | 1.022 |
| | Sev | SC (I) | +|0.019 | 0.00 | 0.00 | -0.281 | 0.876 | 1.022 | 1.023 |
| | Sev1 | SC (I) | +0.001 | 0.00 | 0.00 | -0.216 | 1.963 | 1.022 | 1.023 |
| | SSev | SC (I) | +0.022 | 0.00 | 0.00 | -0.274 | 0.913 | 1.022 | 1.022 |
| | SSev1 | SC (I) | -0.004 | 0.00 | 0.00 | -0.177 | 2.569 | 1.022 | 1.023 |
| NO$_2$ | P | SC (I) | +0.137 | 0.00 | 0.99 | -0.252 | 2.502 | 1.213 | 1.226 |
| | P1 | SC (I) | +0.094 | 0.00 | 1.02 | -0.204 | 2.479 | 1.213 | 1.223 |
| | Mov | SC (D) | +0.196 | 0.01 | 0.99 | -0.314 | 2.351 | 1.213 | 1.232 |
| | Mov1 | SC (D) | +0.185 | 0.01 | 1.00 | -0.266 | 2.195 | 1.213 | 1.225 |
| | Sev | SC (I) | +1.027 | 0.00 | 1.00 | -3.360 | 1.776 | 1.213 | 1.161 |
| | Sev1 | SC (I) | +0.243 | 0.00 | 1.00 | -0.288 | 1.518 | 1.213 | 1.231 |
| | SSev | SC (I) | +1.077 | 0.00 | 1.10 | -3.404 | 1.364 | 1.213 | 1.160 |
| | SSev1 | SC (I) | +0.257 | 0.00 | 1.04 | -0.273 | 1.534 | 1.213 | 1.232 |
| NO | P | SC (D) | +0.010 | 0.00 | 1.00 | -0.089 | 2.748 | 1.166 | 1.165 |
| | P1 | SC (D) | +0.006 | 0.00 | 1.00 | -0.117 | 2.624 | 1.166 | 1.163 |
| | Mov | SC (D) | -0.056 | 0.01 | 0.89 | -0.164 | 2.285 | 1.166 | 1.152 |
| | Mov1 | SC (D) | +0.038 | 0.01 | 0.73 | -0.435 | 1.526 | 1.166 | 1.164 |
| | Sev | SC (I) | +0.059 | 0.00 | 1.07 | -0.219 | 1.462 | 1.166 | 1.160 |
| | Sev1 | SC (I) | +0.915 | 0.00 | 1.12 | -2.788 | 0.000 | 1.166 | 1.273 |
| | SSev | SC (I) | +0.027 | 0.00 | 0.07 | -0.083 | 1.666 | 1.166 | 1.161 |
| | SSev1 | SC (I) | +1.058 | 0.00 | 1.01 | -2.894 | 0.000 | 1.166 | 1.292 |

**Hydrogen sulfide ($H_2S$):**

The initial structure of $H_2S$ molecule is optimized showing $C_{2V}$ symmetry with ~ 1.353Å H-S bond length. We have considered two types of configuration of $H_2S$ molecule to adsorb on the considered Janus monolayers, as shown in Fig.4. The $H_2S$ molecule is first placed on the hollow site of the hexagon for pristine monolayer and allowed adsorption from hydrogen and sulfur site, respectively. In case of a defect included MoSSe monolayer, the gas molecule is placed on the top of the defect site from hydrogen and sulfur sites independently. The gas molecule interaction with H atom side is named with pristine or respective defect, whereas interaction with S atom side is named as pristine or respective defect with 1 as postffix hereafter. The optimized configurations of $H_2S$ molecule on pristine and defect included MoSSe monolayers are shown in Fig.5. We find that MoSSe monolayer transfers small charge ~ 0.011e to $H_2S$ molecule in pristine configuration (P), which acts as an acceptor. In contrast, for pristine1 (P1) configuration, MoSSe monolayer takes small charge ~ 0.001e charge from the $H_2S$ molecule, as shown in Fig.5. The vertical heights for P and P1 configurations are 2.292 Å and 3.019 Å, respectively. Thus, vertical height for P configuration is lower than P1 configuration, which is attributed to the attraction of hydrogen atom in toxic gases to the sulfur atoms in MoSSe monolayer because of their opposite charges. Moreover, in P1 configuration sulfur atom of $H_2S$ molecule is on the top of the selenium atom of MoSSe monolayer that's why the vertical height is relatively more than P configuration. The corresponding adsorption energies are -0.156eV and -0.147eV for P and P1 configurations, respectively. The computed adsorption energy for $H_2S$ molecule is higher than $MoS_2$ (-0.12eV) [9] but less than $WS_2$ monolayer (-0.18eV) [52] monolayer. Fig. 13(a) illustrates that the relative change in adsorption energy is not significant for different $H_2S$ configurations (P and P1). The small adsorption energy signifies the fast recovery of the gas sensor devices.

We further analyzed the electronic properties for all the considered configuration of $H_2S$ adsorption. The spin-polarized band structures for all considered configuration are plotted and compared with and without adsorbed gas molecule, and shown in Fig. S1 (See supplementary data). Fig.S1 (a) shows the band structure of the pristine and H$_2$S adsorbed molecule in two different configurations P and P1 for MoSSe monolayer. We do not notice any additional peak near to the Fermi energy in the band structure as compared to without H$_2$S molecule adsorbed MoSSe monolayer. The band gap is also not changing significantly due to H$_2$S adsorption and the noticed symmetry between spin up and spin down states confirms the non-magnetic semiconducting behavior of MoSSe monolayer even after adsorbing H$_2$S molecule. Defect included monolayer-like Mo$_V$, Se$_V$ and S/Se$_V$ are also considered to see the impact of defects on the adsorption behavior. The optimized configurations of H$_2$S molecule adsorbed on the defect included monolayers and corresponding charge transfer are shown in Fig 5. The value of charge transfer analyzed using the Bader charge analysis is listed in the respective Figs (5). The variation in adsorption energy of H$_2$S molecule against these various configurations of the defect included monolayers is summarized in Fig.14 (a), showing the maximum adsorption energy ~ -0.238 eV and corresponding minimum vertical height ~ 1.609 Å for SSe$_V$ configuration. The vertical height of H$_2$S molecule versus different defect included monolayers is plotted in Fig.14 (b). The noticed low values of adsorption energies with relatively larger vertical heights support the physisorption between the adsorbent H$_2$S and pristine or defect included MoSSe host material.

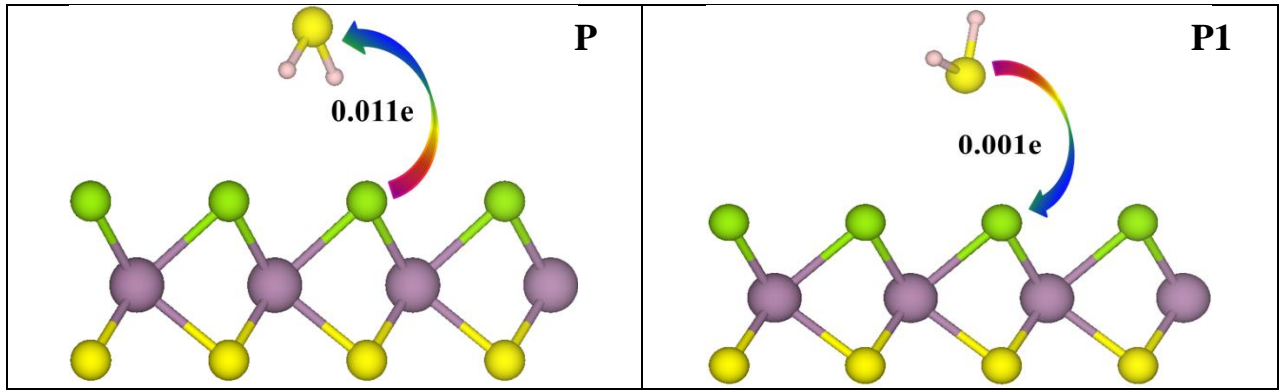

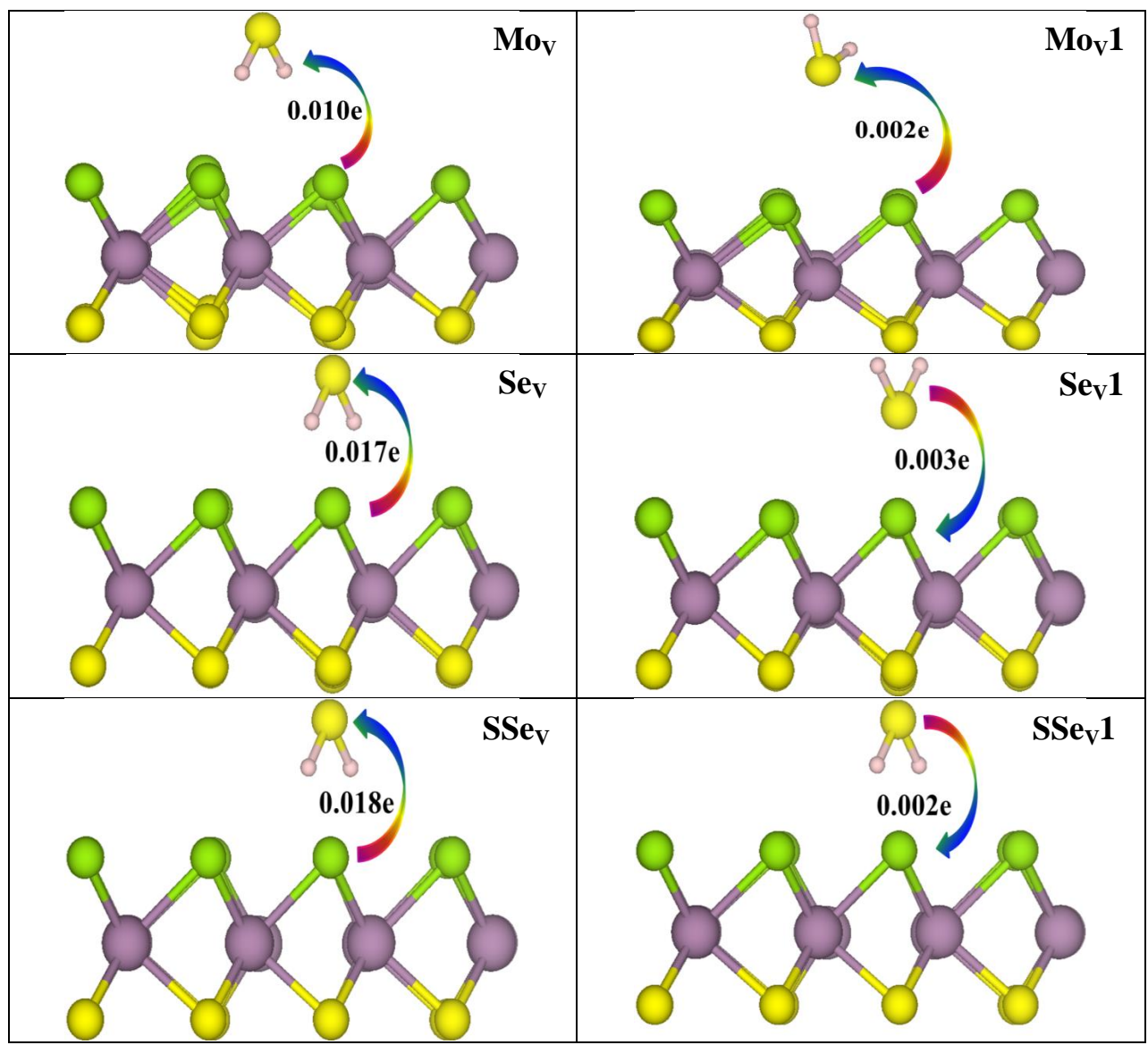

Fig.5 Optimized geometry and respective charge transfer for the adsorbed $H_2S$ gas molecule on the pristine, molybdenum, selenium, and sulfur/selenium defect included Janus MoSSe monolayer. [Pink, yellow, green, and purple represent the hydrogen, sulfur, selenium, and molybdenum atom respectively]

Thus, we find that the most stable configuration for the adsorption of $H_2S$ molecule is $Se_V$ and $S/Se_V$ configurations in terms of the vertical height, adsorption energy, and charge sharing. In $Se_V$ and $S/Se_V$ configuration, MoSSe monolayer transfers about 0.017e and 0.018e charge to $H_2S$ molecule, respectively because of dangling bonds in Mo atom in MoSSe monolayer due to the $S/Se_V$ vacancies. Thus, the charge transfer from MoSSe monolayer to $H_2S$ molecule will increase the resistance and the Mo dangling bonds, having excess positive charge will repel hydrogen atom, resulting in an elongated H-S bond length by 0.002Å, whereas attracting the sulfur near to vacancy site, reducing the vertical height. These studies support the vdW interaction for $H_2S$ molecule, confirming the physisorption between the adsorbent and host. The band structure of defect included MoSSe monolayer along with adsorbed $H_2S$ molecule are shown in Fig.S1, confirming the contribution of bands due to $H_2S$ molecule are overlapping with defect included MoSSe Janus monolayer and also do not show any addition bands into the forbidden region. Moreover, the degeneracy of the band is broken in case of Mo defect included MoSSe layer with $H_2S$ adsorbed molecule. The CDD plots for an $H_2S$ molecule with pristine and defect included monolayers are plotted in the Fig. S2 and the noticed charge sharing is in agreement with Bader charge analysis. The smaller interactions with respective charge analysis support the vdW interaction between adsorbate and host. Thus, these studies suggest that defects assist in improving the adsorption of the $H_2S$ gas molecule and the maximum for $S/Se_V$ defect. The relatively low adsorption energy of $H_2S$ molecule for pristine and defect included MoSSe monolayer will lead to fast recovery time and thus, may be suitable for fast sensing.

**Ammonia ($NH_3$):**

The initial structure of ammonia molecule is optimized and observed $C_{3V}$ symmetry with ~1.022 Å N-H bond length. The adsorption behavior of ammonia gas molecule on the various configurations of the monolayer-like the pristine and defect included monolayers are investigated. Here, also two configurations of $NH_3$ molecule are considered, as shown in Fig.4. The first one is hydrogen atom pointing toward the host monolayer and another is nitrogen atom pointing towards the host monolayer. The optimized structure of P configuration shows about 0.005e charge transfer from $NH_3$ to the host monolayer and thus, behaving as a donor. The similar behavior is also observed for P1 configuration, where about 0.028e charge transfer is noticed from $NH_3$ molecule to the monolayer. The Fermi level of MoSSe monolayer lies between the highest occupied molecular orbital (HOMO) and lowest occupied molecular orbital (LUMO) of $NH_3$, and this suitable position of energy levels is the main source of noticed charge transfer from the $NH_3$ molecule to monolayer [53]. This is consistent with other reports showing the adsorption of $NH_3$ on $MoS_2$ and $WS_2$ monolayer giving rise to the donor behavior, whereas the adsorption energy is higher for MoSSe Janus monolayer than the previous reports [9,17,19,53,54]. Adsorption energies and vertical heights for P and P1 configurations are shown in Fig.13, confirming the physisorption or vdW interaction between adsorbent and host. These studies suggest that P1 configuration is relatively more stable as compared to P configuration for $NH_3$ adsorption because of large adsorption energy and charge transfer for P1 configuration. Huo et al. experimentally demonstrated the sensing of $NH_3$ using $WS_2$ - FET (Field Effect Transistor) device, discussing nearly the similar sensing mechanism [55]. Here, $WS_2$ is n-type semiconductor, with intrinsic electrons present in $WS_2$ –FET channel. The channel electron density showed enhancement after exposing to $NH_3$, suggesting the electron transfer from $NH_3$ to the host $WS_2$ layer i.e. channel [50]. The computed spin polarized band structure of $NH_3$ adsorbed P and P1 configurations are shown in Fig.S3 (see supplementary data). The band structure shows no significant change after the adsorption of $NH_3$ molecule in the forbidden region and also spin up and spin down states are highly symmetric, confirming the non-magnetic semiconductor behavior even after adsorption of $NH_3$ molecule. No change in band structure and thus effectively no change in their respective conductivity after adsorbing $NH_3$ molecule to the host MoSSe Janus layer. The CDD plots for P and P1 configurations are shown in Fig.S4 (see supplementary data), validating the Bader charge analysis. This CDD is supporting the physisorption between the adsorbent $NH_3$ and host MoSSe Janus monolayer. There are reports suggesting that the defects included TMDCs based monolayer improve the sensitivity and selectivity of gases [56]. Further, we considered $Mo_V$, $Se_V$ and $S/Se_V$ defects in Janus MoSSe monolayer to understand their impact on the adsorption behavior of $NH_3$ molecule. The optimized configurations for $NH_3$ adsorbed defect included MoSSe monolayer in conjunction with respective charge sharing are shown in Fig.6. The vertical height between the adsorbent and host material confirm the physisorption between $NH_3$ and host layer, listed in Table1. $Mo_V$ defect included Janus monolayer creates the unsaturated bonds nearby sulfur and selenium atoms. $NH_3$ adsorbed $Mo_V$ configuration shows that $NH_3$ molecule transfer about 0.010e charge to the monolayer and a similar behavior is observed for $Mo_V1$ configuration with much smaller charge transfer 0.004e to monolayer and thus, acting as a donor. Adsorption energy of $Mo_V$ and $Mo_V1$ configurations are -0.146eV and -0.160eV, respectively which are in the order or less than that of the pristine configurations. $Mo_V$ defect included Janus MoSSe monolayer does not exhibit any significant improvement in the $NH_3$ adsorption. The spin-polarized band structure for $Mo_V$ and $Mo_V1$ configurations are shown in Fig.S3 (b) (see supplementary data), confirming that no significant changes are observed after $NH_3$ adsorption. The spin up and spin down states maintain their symmetry in the band structure, showing the non-magnetic semiconducting behavior. Next adsorption configuration is $NH_3$ adsorbed $Se_V$ defect included MoSSe monolayer. The optimized structure for $NH_3$ adsorbed $Se_V$ and $Se_V1$ configurations are shown in Fig.6 and found that monolayer transfers about 0.019e and 0.001e charge to $NH_3$ molecule in $Se_V$ and $Se_V1$ configurations and thus, acting as an acceptor. Adsorption energies (vertical heights) for $Se_V$ and $Se_V1$ configurations are -0.281eV (0.876Å) and -0.216eV (1.963Å), respectively and also summarized in Fig.14. These values of adsorption energies and vertical heights confirm the physisorption between the monolayer and $NH_3$ molecule. The relatively large value of adsorption energy in case of $Se_V$ defect included Janus MoSSe monolayer suggests the improvement in $NH_3$ adsorption with respect to pristine and other defect included MoSSe monolayer. In this case, the electrical conductivity of MoSSe monolayer may decrease due to noticed change transfer from the monolayer to $NH_3$ molecule. These moderate values of adsorption energy will lead to the smaller recovery time.

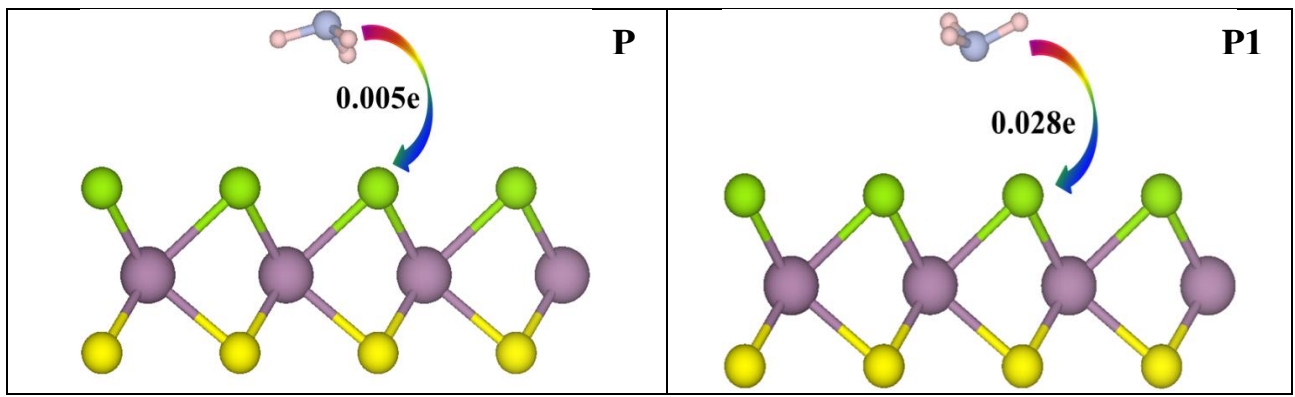

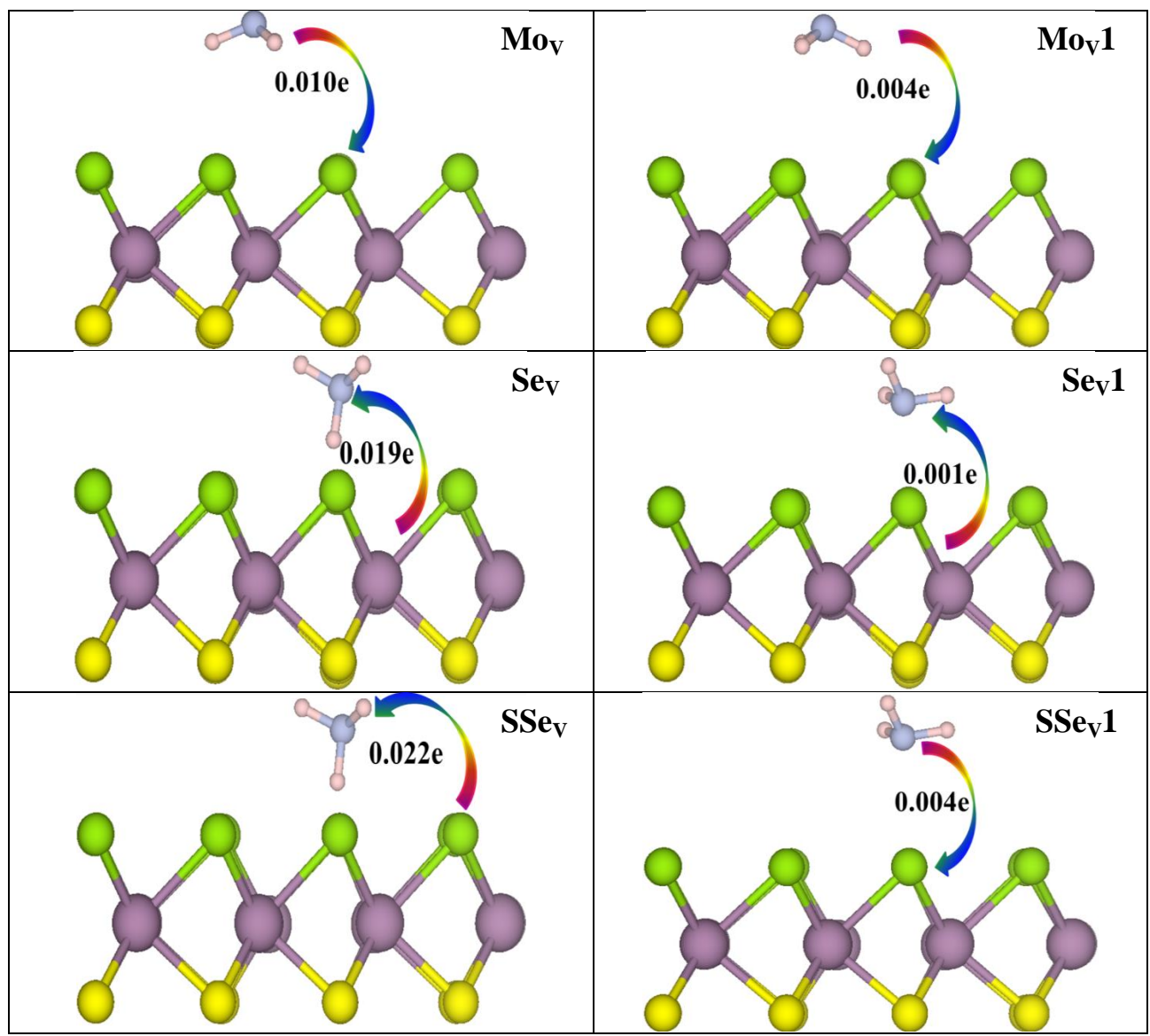

Fig.6 Optimized geometry and corresponding charge transfer for the adsorbed NH$_3$ gas molecule on the pristine, molybdenum, selenium, and sulfur/selenium defect included Janus MoSSe monolayer. [Grey, pink, green, purple, and yellow represents the nitrogen, hydrogen, selenium, molybdenum, and sulfur respectively]

The other possible adsorption configurations are SSe$_V$ and SSe$_V$1 for NH$_3$ molecule and the respective optimized geometries are shown in Fig 6. The charge of about 0.022e and 0.004e is transferred from the monolayer to NH$_3$ molecule for SSe$_V$ and SSe$_V$1 configurations, respectively. Adsorption energies (vertical heights) for SSe$_V$ and SSe$_V$1 configurations are -0.274eV (0.913Å) and -0.177eV (2.569), respectively. Thus, in this case too, NH$_3$ molecules are interacting through the weak vdW interactions, confirming the physisorption, similar to the other considered configurations. Adsorption energy for SSe$_V$ configuration is -0.274eV, the highest for NH$_3$ molecule among the considered configurations. The spin polarized band structure and density of states suggest no significant change in their electronic properties and the symmetry of spin up and down density of states confirm the non-magnetic semiconducting behavior of NH$_3$ adsorbed MoSSe monolayer with S/Se$_V$ defect. The above study suggests that the defect is a prominent source to improve the adsorption energy. The adsorption energy is related to the recovery time, so due to small adsorption energy the recovery time also will be small. The CDD for NH$_3$ adsorbed molecule on S/Se$_V$ defect included monolayers are shown in Fig.S4. The noticed charge transfer between the adsorbent and host is relatively small and thus supporting vdW interaction between NH$_3$ and host layer causing the physisorption.

**Nitrogen dioxide (NO$_2$):**

The initial structure of nitrogen dioxide is optimized and found C$_{2V}$ point group symmetry with 1.21 Å optimized N-O bond length. Here we considered two configurations of the NO$_2$ molecule for adsorption studies as shown in Fig.4. In the first case, oxygen atom is pointing towards MoSSe monolayer and in another case, nitrogen atom is pointing towards MoSSe monolayer. Pristine, Mo$_V$, Se$_V$ and S/Se$_V$ defect included MoSSe Janus monolayers are considered as a host to understand the adsorption of NO$_2$ molecule. The optimized geometry of NO$_2$ adsorbed with MoSSe layer in P and P1 configurations are shown in Fig.7. A charge transfer of about 0.137e and 0.094e charge is noticed from pristine MoSSe monolayer to NO$_2$ molecule in P and P1 configurations, respectively. This is attributed to the deficiency of electron in NO$_2$ molecule, and thus, the acceptor behavior of NO$_2$ molecule can be inferred. Further, the lowest unoccupied molecular orbital (LUMO) of NO$_2$ lies below the Fermi level of Janus MoSSe monolayer, which may also favor the electron transfer from MoSSe monolayer to NO$_2$ molecule [53]. A similar behavior is noticed for NO$_2$ adsorption in MoS$_2$ and WS$_2$ monolayers [53,54]. Adsorption energies of NO$_2$ molecule for P and P1 configurations are -0.252eV and -0.204eV, respectively, whereas the vertical heights are nearly same for both the configuration, as summarized in Fig.13. The relatively smaller values of adsorption energies and large vertical heights confirm the physisorption. The computed value of adsorption energy is larger than the previously reported value under the same level of theory (DFT-D2) [8]. The N-O bond length is elongated due to adsorption and changed from 1.21Å to 1.22Å after adsorption. The spin-polarized band structure for $NO_2$ adsorbed pristine Janus MoSSe monolayers are shown in Fig.9 and compared with a pristine Janus MoSSe monolayer without adsorbed $NO_2$ molecule. We noticed the p-type semiconducting behavior for $NO_2$ adsorbed MoSSe monolayer. This is because of the localized states originating from the adsorption of $NO_2$ molecule just above the valence band, showing no hybridization between the adsorbent and host material. A similar behavior is also predicted by Zhou et al. [53] for the adsorption of $NO_2$ on $WS_2$ monolayer. Further, the noticed spin splitting in the band structure supports the onset of the finite magnetic moment for $NO_2$ adsorbed monolayer. The magnetic moments for P and P1 configurations are 0.99 $\mu_B$ and 1.02 $\mu_B$ for $NO_2$ adsorbed MoSSe Janus monolayer, respectively, These results are consistent with results reported by Sahoo et al for $NO_2$ adsorbed $MoS_2$ monolayer [18]. The CDD plots for P and P1 configurations are shown in Fig.7, confirming that $NO_2$ molecule accumulates the charge, consistent with the Bader charge analysis. Charge accumulation in P configuration is a bit more than P1 configuration. These results for $NO_2$ adsorption on the pristine Janus MoSSe monolayer suggest that the adsorption energy is larger than that of reported for $MoS_2$ and $WS_2$ monolayers.

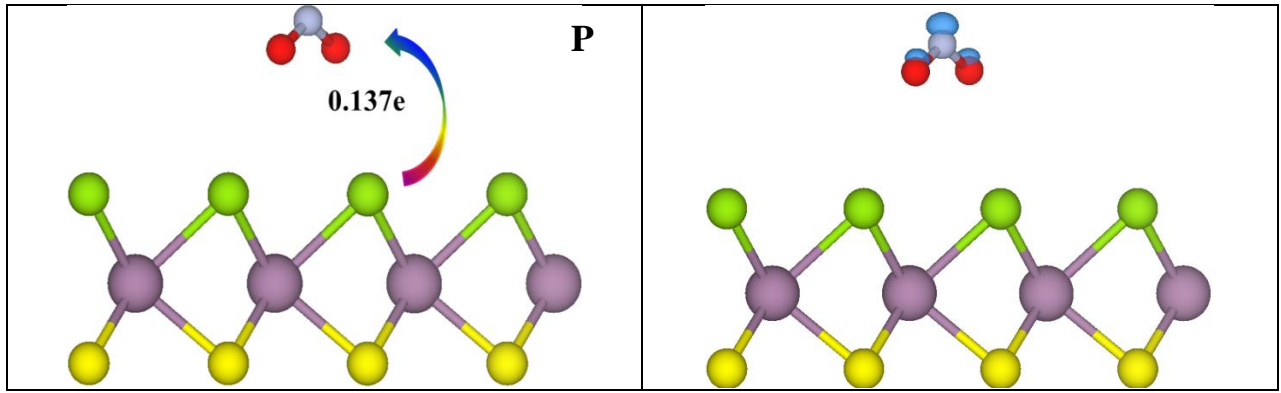

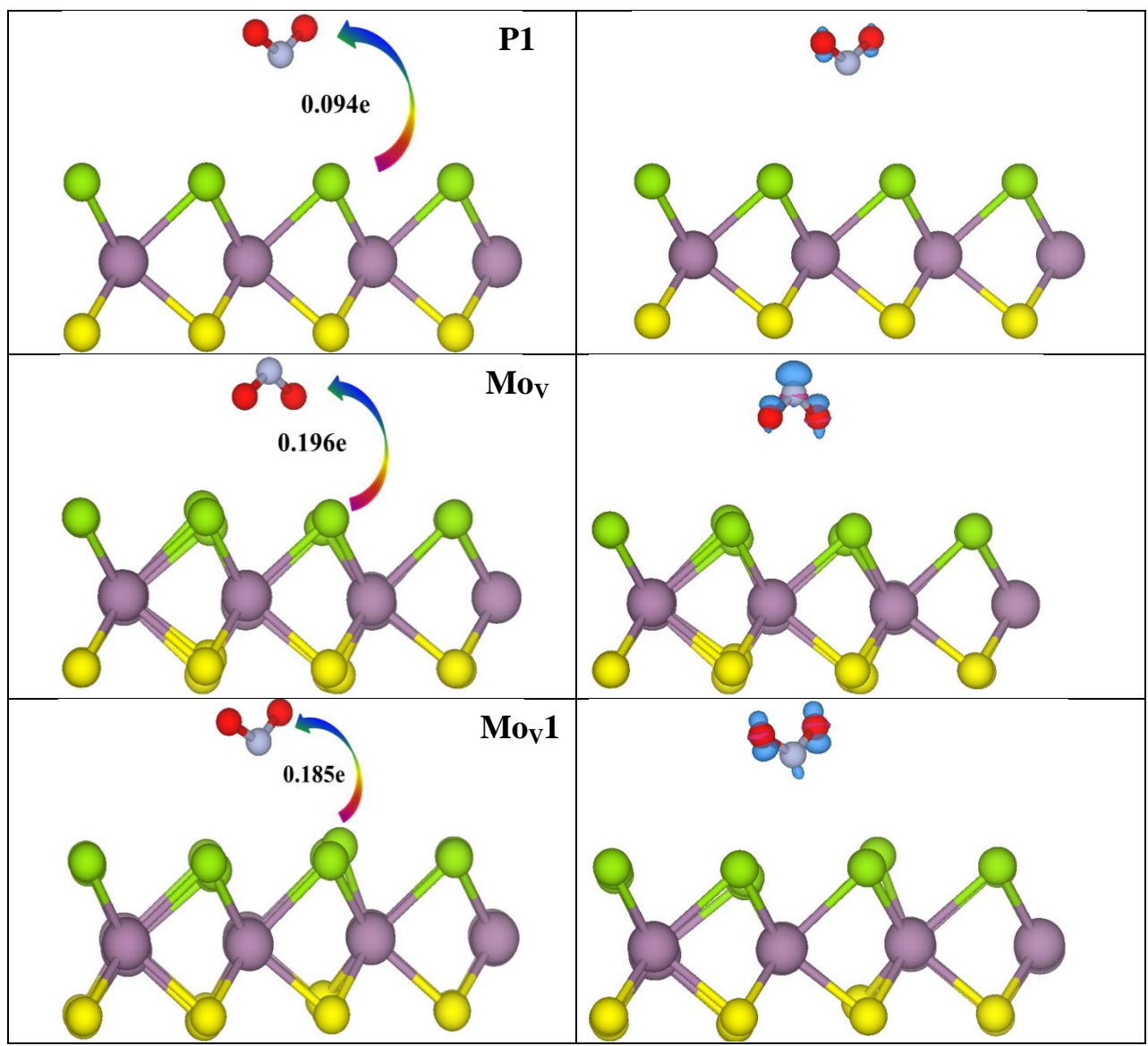

Fig.7 Optimized geometry included charge transfer, and respective charge difference density in the adsorption of NO$_2$ gas molecule on the pristine and molybdenum defect included Janus MoSSe monolayer. [Grey, red, green, purple and yellow represents the nitrogen, oxygen, selenium, molybdenum, and sulfur, respectively]

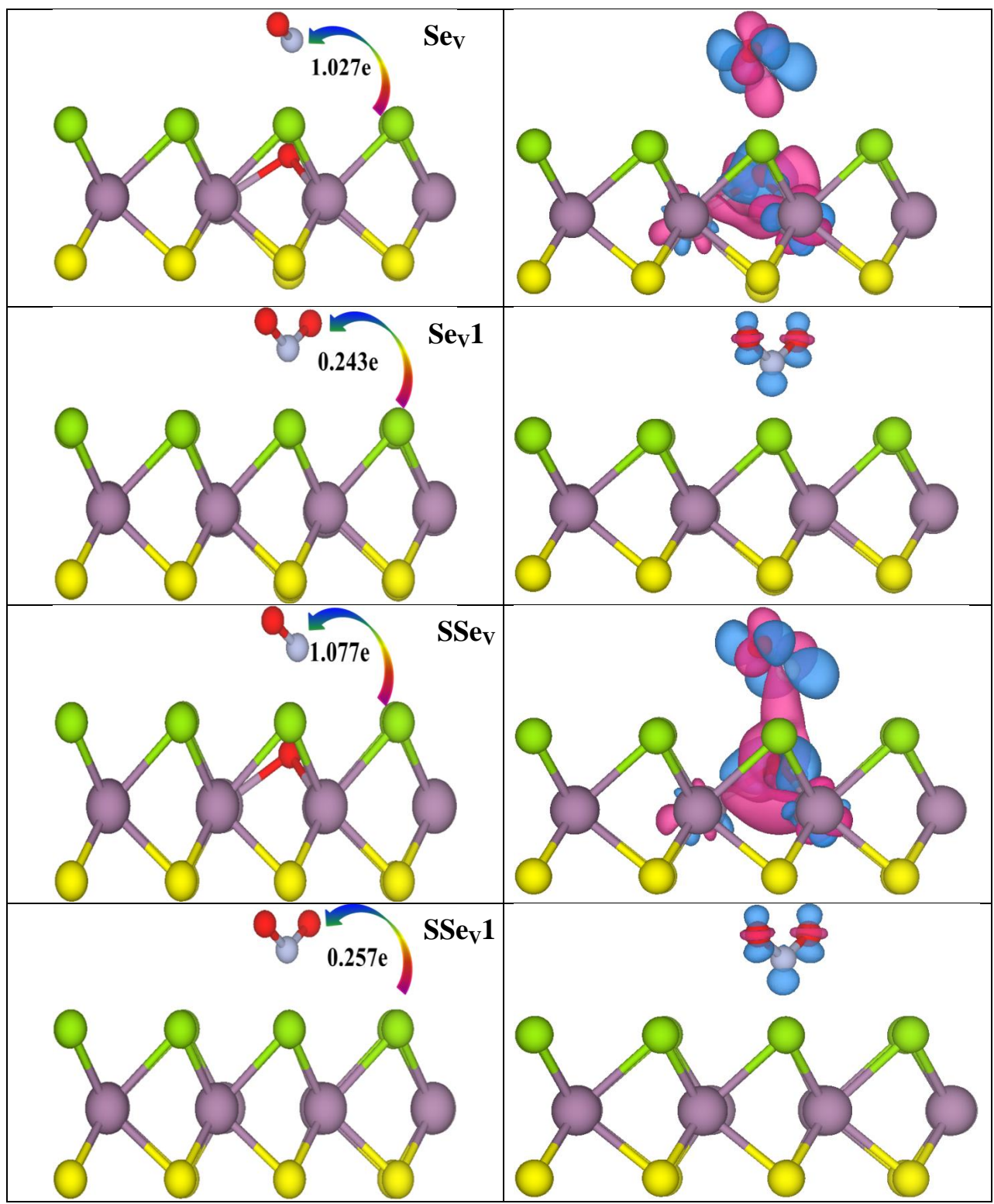

Fig.8 Optimized geometry included charge transfer, and corresponding charge difference density in the adsorption of $NO_2$ gas molecule on the selenium and sulfur/selenium defect included Janus MoSSe monolayer. [Grey, red, green, purple and yellow represents the nitrogen, oxygen, selenium, molybdenum and sulfur, respectively]

Further, the optimized geometry of $Mo_V$ and $Mo_V1$ configuration are shown in Fig.7 and supporting the acceptor behavior of $NO_2$ molecule as about 0.196e and 0.185e charge transfer is noticed from monolayer in $Mo_V$ and $Mo_V1$ configurations, respectively. Thus, the charge density of $Mo_V$ defect included monolayer is reduced under the exposer of $NO_2$ gas molecule, which may result in enhanced resistance of MoSSe monolayer. Adsorption energies for $Mo_V$ and $Mo_V1$ configuration are -0.314eV and 0.266eV respectively and corresponding vertical heights are shown in Fig.14 (b), supporting the physisorption. Due to the adsorption, the bond length of N-O is changed from 1.21 Å to 1.23 Å and 1.22 Å for $Mo_V$ and $Mo_V1$ configurations, respectively. The CDD plots for $Mo_V$ and $Mo_V1$ configurations show the more charge accumulation at $NO_2$ molecule site, in agreement with Bader charge analysis. The spin polarized band structures for $NO_2$ adsorbed $Mo_V$ and $Mo_V1$ configurations are shown in Fig.9 (b) and also compared with pristine Janus MoSSe monolayer. We noticed the onset of electronic states near the Fermi energy, showing spin splitting causing the onset of the total non-zero magnetic moment after $NO_2$ adsorption in conjunction with semiconducting nature. The noticed magnetic moment is about $1.00\mu_B$ for both $Mo_V$ and $Mo_V1$ configurations. A significant change in the bond length is also observed and respective changes are listed in Table1. So, $Mo_V$ defect included monolayer significantly improves the adsorption behavior of $NO_2$ molecule as compared to the pristine monolayer. The optimized geometry of $Se_V$ defect included $Se_V$ and $Se_V1$ configurations are shown in Fig.8. In $Se_V$ configuration, oxygen atoms are pointed towards the monolayer, which dissociated after optimization of $NO_2$ molecule, and forming the oxygen atom doped Janus MoSSe monolayer and resembling like NO molecule adsorption with 1.161Å bond length, closes to the isolated NO molecule. The adsorption energy for $NO_2$ molecule in $Se_V$ configuration is found to be -3.360eV. Further, MoSSe monolayer in $Se_V$ configuration transfers 1.027e charge to the NO molecule and act as an acceptor. A similar behavior has also been observed for $NO_2$ molecular adsorbed WSe$_2$ monolayer with Se defects [13]. The spin polarized band structure of MoSSe monolayer with Se$_V$ configuration is shown in Fig.9 (c) and observed the selenium vacancy defect states are vanished due to the dissociation of an oxygen atom at the defect site. It also gives rise to the spin splitting in valence and conduction bands, confirming the onset of magnetic behavior with 1$\mu_B$ magnetic moment. The optimized geometry for Se$_V$1 configuration is shown in Fig.8 and observed that 0.243e charge is transferred from the monolayer to NO$_2$ molecule, which confirms the physisorption. The adsorption energy (vertical height) for Se$_V$1 configuration is -0.288eV (1.518Å) and the optimized bond length of NO$_2$ molecule is elongated by 0.02Å due to the charge transfer. CDD plots for Se$_V$ and Se$_V$1 configurations are shown in Fig.8, illustrating that more charge sharing between the adsorbent and host material in Se$_V$ configuration as compared to that of Se$_V$1 configuration. These results are also validated using the Bader charge analysis. Se$_V$ defect included monolayer improves the adsorption energy as compared to the pristine and Mo$_V$ defect included monolayers. The improvement in the adsorption energy leads to enhance the recovery time of the gas sensor device. Adsorption effect of NO$_2$ molecule is also investigated for S/Se$_V$ defect included MoSSe monolayer and optimized SSe$_V$ and SSe$_V$1 configurations are shown in Fig.8. SSe$_V$ configuration shows the similar behavior like Se$_V$ configuration, where NO$_2$ molecule dissociates and forming the oxygen doped MoSSe monolayer resembling like NO molecule adsorption. Adsorption energy for adsorption of NO$_2$ molecule in SSe$_V$ configuration is -3.404eV, higher than the Se$_V$ configuration, as summarized in Fig.14. The bond length of adsorbed NO molecule is 1.160 Å comparable to that of isolated NO molecule. The spin polarized band structure for SSe$_V$ configuration is shown in Fig.9 (d), showing the additional electronic states as compared to S/Se$_V$ defect included monolayer due to the addition 1.077e charge depletion in monolayer and spin splitting induced 1.10$\mu_B$ magnetic moment. Adsorption energy and vertical height for SSe$_V$1 configuration are -0.273eV and 1.534Å, respectively. Here also NO$_2$ molecule accumulates excess 0.257e charge from monolayer and acts as an acceptor. The spin polarized band structure of SSe$_V$1 configuration is shown in Fig.9 (d), exhibiting the additional electronic states above the Fermi energy in the conduction band due to the charge transfer from a monolayer to NO$_2$ molecule. The CDD plots of SSe$_V$ and SSe$_V$1 configurations are shown in Fig.8, in agreement with the Bader charge analysis.

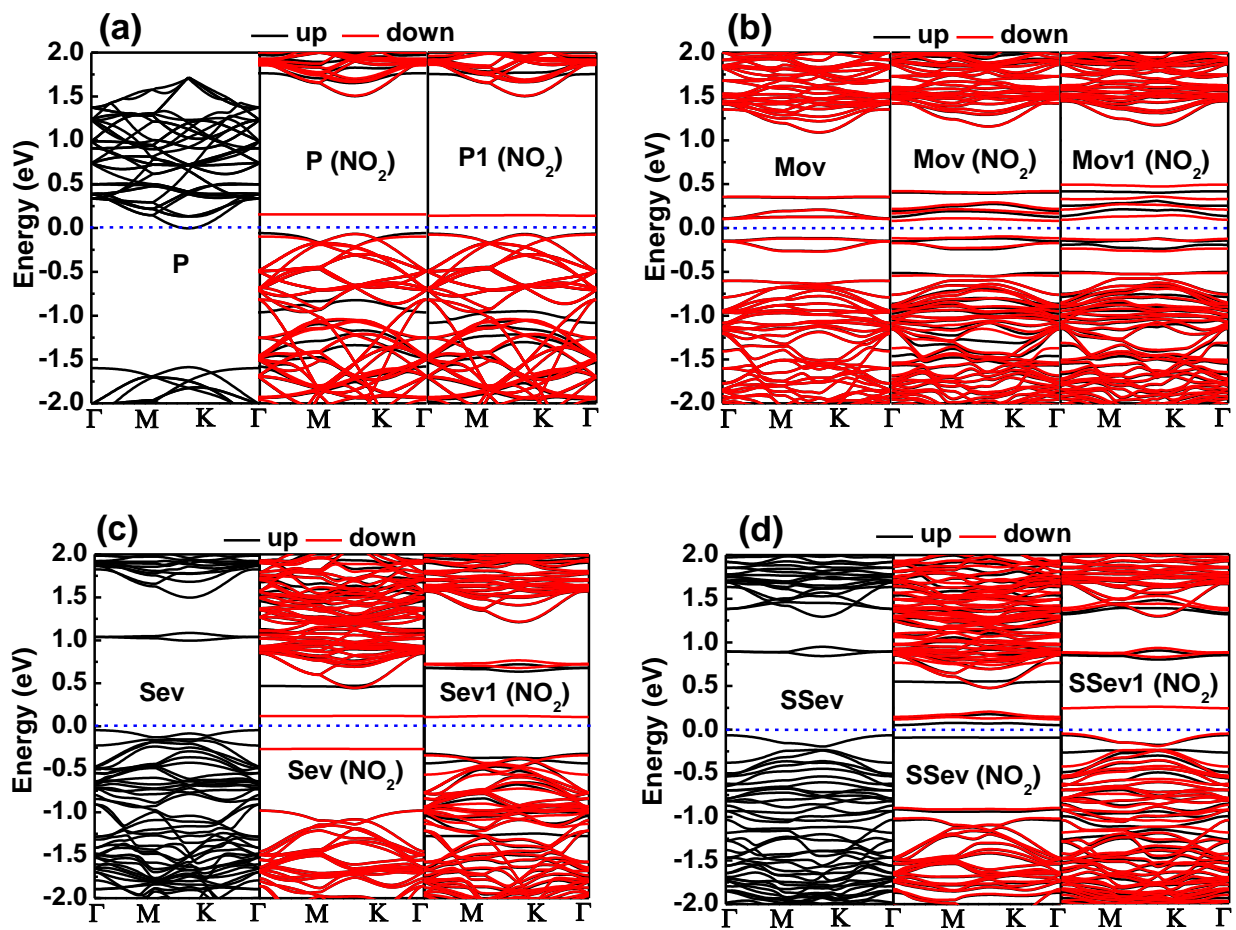

Fig.9 Spin-polarized band structure for (a) pristine monolayer (b) Molybdenum defect (c) selenium defect and (d) sulfur/selenium included monolayer along with NO$_2$ molecule adsorption. [Black and red represents the spin up and spin down states respectively]

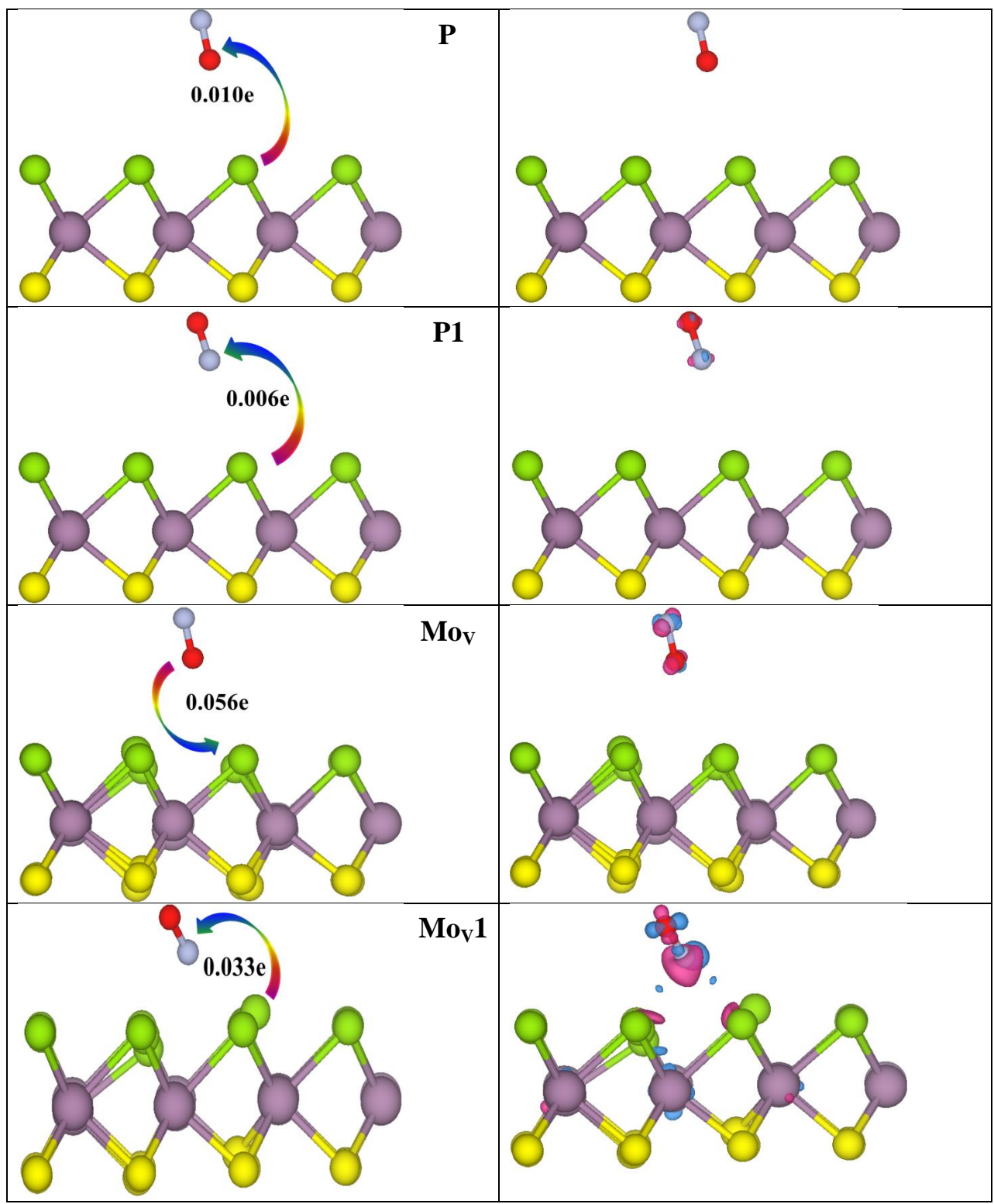

Fig.10 Optimized geometry included charge transfer, and respective charge difference density in the adsorption of NO gas molecule on the pristine and molybdenum defect included Janus MoSSe monolayer. [Grey, red, green, purple and yellow represents the nitrogen, oxygen, selenium, molybdenum and sulfur respectively]

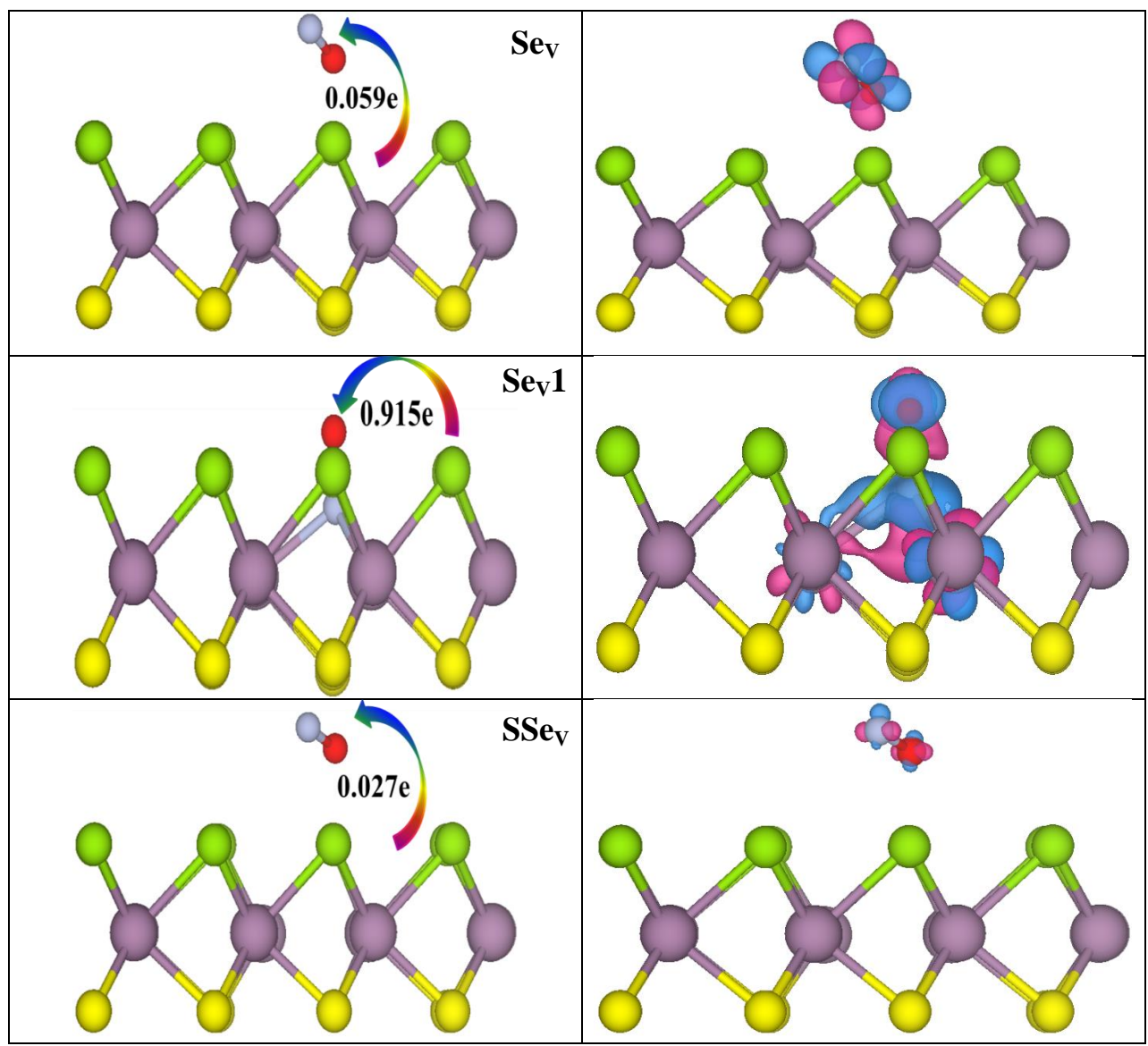

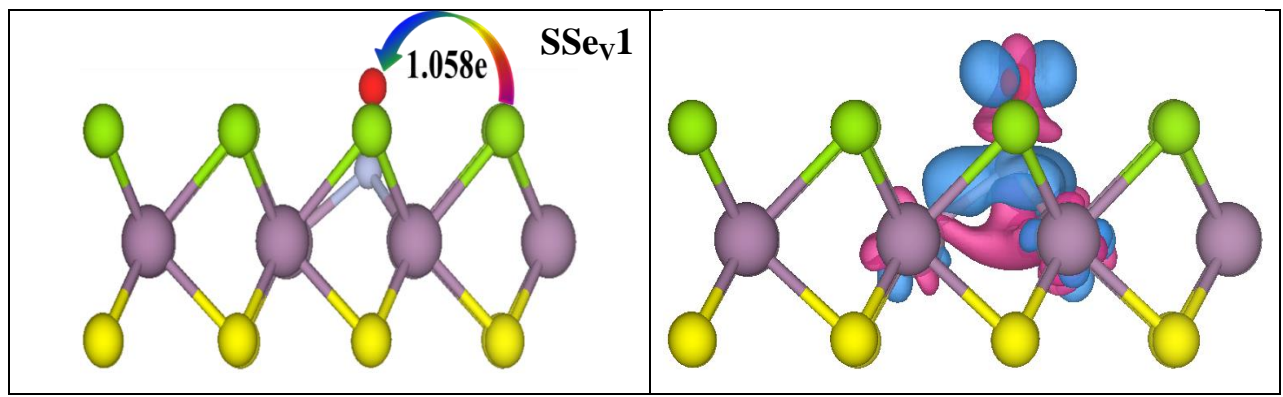

Fig.11 Optimized geometry included charge transfer, and corresponding charge difference density in the adsorption of NO gas molecule on the selenium and sulfur/selenium defect included Janus MoSSe monolayer. [Grey, red, green, purple and yellow represents the nitrogen, oxygen, selenium, molybdenum, and sulfur respectively]

**Nitric oxide (NO):**

Nitric oxide is a linear molecule and has the point group symmetry $C_{\infty V}$. Nitric oxide behaves like a free radical because of its one unpaired electron. Adsorption of NO molecule is evaluated in two configurations (i) oxygen is pointing towards to the monolayer and (ii) nitrogen atom is pointing towards to the monolayer. Pristine and defect included monolayers are considered as a host material to see the adsorption of NO molecule. Pristine Janus MoSSe monolayer is considered to adsorb the NO molecule. The optimized structures for P and P1 configurations are shown in Fig 10, where a charge transfer of 0.010e and 0.006e took place from a monolayer to NO molecule in P and P1 configurations, showing the acceptor nature of NO molecule. The possible reasons for charge transfer from a monolayer to NO molecule may be the position of the lowest unoccupied molecular orbital (LUMO) of NO molecule, which lies below the Fermi level of Janus MoSSe monolayer [53]. Vertical heights for P and P1 configurations are 2.748Å and 2.624Å whereas the adsorption energies are -0.089eV and -0.117eV, respectively. The similar adsorption behavior also has been reported by Yue et al. [19] for NO adsorbed $MoS_2$ monolayer. The spin polarized band structure for P and P1 configurations are shown in Fig 12 (a), and compared with pristine MoSSe monolayer which illustrates the onset of electronic states above and below the Fermi energy. This is in agreement with NO adsorbed MoS$_2$ monolayer results [19]. Further, the spin splitting in the spin-polarized band structure of P and P1 configurations also showed the onset of 1μ$_B$ magnetic moment, which is equal to that of NO adsorbed in MoS$_2$ monolayer [16]. These states are coming due to the charge transfer of about 0.010e and 0.006e from the monolayer to the NO adsorbent for P and P1 configurations, respectively, supporting the acceptor behavior of NO molecule. In this case, the bond length of the adsorbed molecule does not change significantly as the charge transfer is relatively smaller as compared to other cases. Janus MoSSe monolayer is an n-type semiconductor which has electron as carriers, however, during NO adsorption a charge transfer from monolayer to NO molecule is taking place, and thus depleting electrons from monolayer and thus converting into a p-type semiconductor [57]. Further, this is substantiated by the noticed VBM shift towards the Fermi energy and CBM away from the Fermi energy, Fig 12 (a). This may increase the monolayer resistance and thus reduce the conductivity. Mo$_V$ defect included monolayer is also considered to understand the adsorption of NO molecule. The optimized NO adsorbed MoSSe monolayer structure for Mo$_V$ and Mo$_V$1 configurations are shown in Fig.10. Adsorption energies (vertical heights) for Mo$_V$ and Mo$_V$1 configurations are -0.164eV (2.285Å) and -0.435eV (1.526Å), respectively. Adsorption energy increased significantly after introducing Mo$_V$ defect in Janus MoSSe monolayer. A charge transfer of about 0.056e from NO molecule to monolayer and 0.033e is noticed from the monolayer to NO absorbent, respectively and corresponding changes in NO bond length are 0.014Å and 0.002Å, respectively. The spin polarized band structure for Mo$_V$ and Mo$_V$1 configurations are shown in Fig.12 (b), confirming the spin splitting, and thus, giving rise to non-zero magnetic moment 0.89μ$_B$ and 0.73 μ$_B$, respectively. Selenium defect included Janus MoSSe monolayer is also investigated for the adsorption of NO molecule. The optimized geometry for Se$_V$ and Se$_V$1 configurations are depicted in Fig.11, confirming that in Se$_V$ configuration, 0.059e charge is transferred from monolayer to NO molecule in Se$_V$ and 0.915e charge to the NO molecule in Se$_V$1 configuration. This charge transfer from the monolayer to NO molecule in Se$_V$ and Se$_V$1 configurations supports the acceptor behavior of NO molecule. The more charge sharing in Se$_V$1 configuration is attributed to formation of a covalent bond between nitrogen and nearby tungsten atom. This bond formation confirms the chemisorption of NO molecule with Janus monolayer in Se$_V$1 configuration in contrast to other observations as discussed earlier. Moreover, Se$_V$ configuration shows the physisorption behavior. The chemisorption of Se$_V$1 configuration and physisorption behavior of Se$_V$ configuration is also supported by CDD plots, shown in Fig.11, consistent with the Bader charge analysis. The spin polarized band structure of Se$_V$ configuration shows the defect states in the VB and CB due to the adsorption of NO molecule. In Se$_V$1 configuration, the selenium vacancy defect states are removed due to the formation of Mo-N covalent bond. The spin-polarized band structure shows the spin splitting for both Se$_V$ and Se$_V$1 configurations, giving rise to non-zero magnetic moment of 1.07μ$_B$ and 1.12 μ$_B$, respectively after NO adsorption. The adsorption energies for Se$_V$ and Se$_V$1 configurations are -0.219eV and -2.788eV, respectively. Thus, Se$_V$ defect included Janus MoSSe monolayer is more prominent to adsorb NO molecule, which may exhibit relatively large recover time.

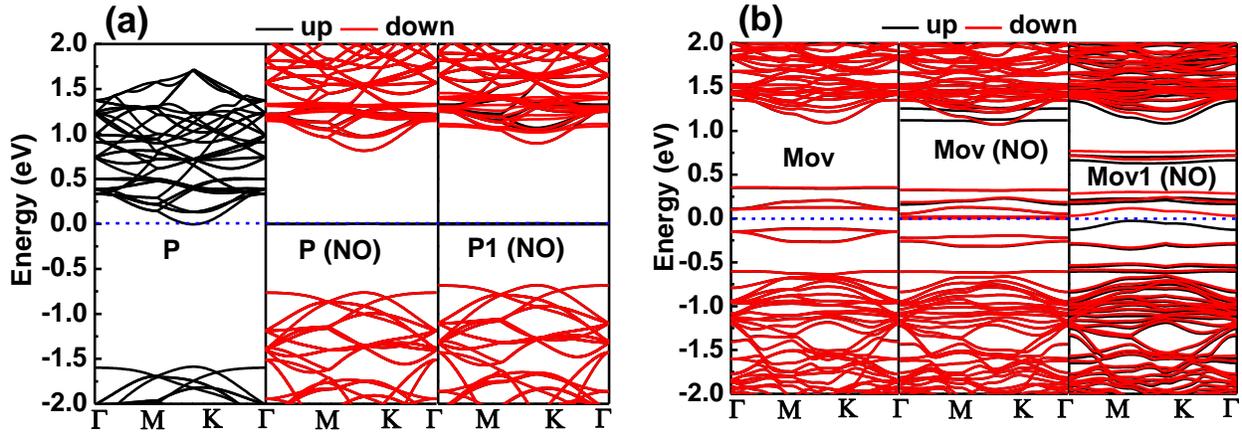

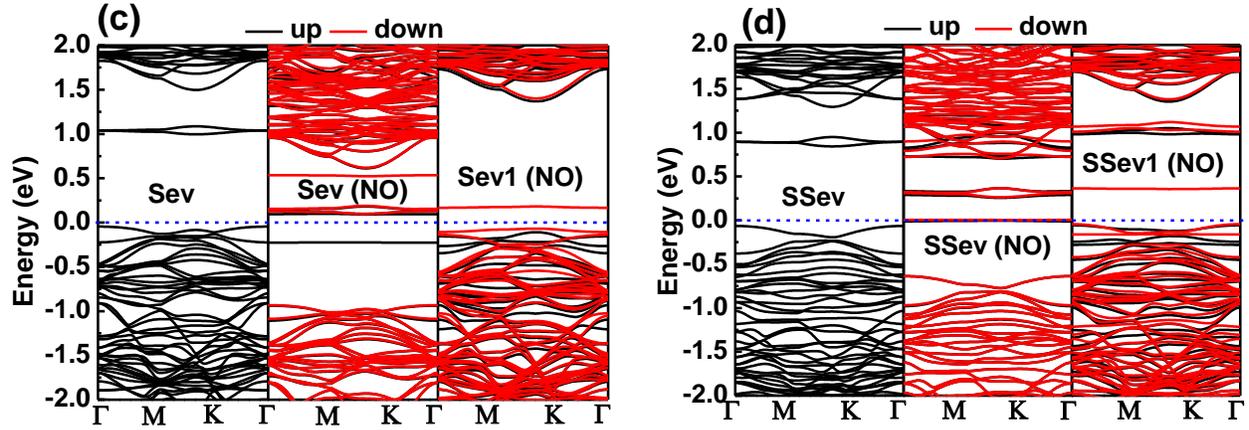

Fig.12 Spin-polarized band structure of (a) pristine monolayer (b) Molybdenum defect (c) selenium defect and (d) sulfur/selenium included monolayer along with NO molecule adsorption. [Black and red represents the spin up and spin down states respectively]

The optimized geometry of $SSe_V$ and $SSe_V1$ configurations are shown in Fig.11, supporting the physisorption of NO molecule for $SSe_V$ configuration where a charge transfer of about 0.027e from a monolayer to NO molecule is noticed. Further, NO molecule accepts about 1.058e charge from monolayer in $SSe_V1$ configuration due to the chemisorption with nearby Mo atom in MoSSe monolayer. Adsorption energy for $SSe_V$ and $SSe_V1$ configurations are -0.083eV and -2.894eV, respectively. Thus, we find that adsorption energy of NO molecule in $SSe_V1$ configuration is much larger than the above considered host materials. The bond length of NO molecule has elongated from 1.161Å to 1.292Å due to tis chemical bonding with nearby molybdenum atoms. CDD plots for $SSe_V$ and $SSe_V1$ configurations are shown in Fig.11, supporting the noticed physisorption of NO molecule in $SSe_V$ configuration and chemisorption in $SSe_V1$ configuration, which is consistent with Bader charge analysis. The spin polarized band structure for $SSe_V$ and $SSe_V1$ configurations are shown in Fig.12 (d) and we noticed the spin splitting in $SSe_V1$ configuration, inducing 1.01$\mu_B$ magnetic moment after NO adsorption.

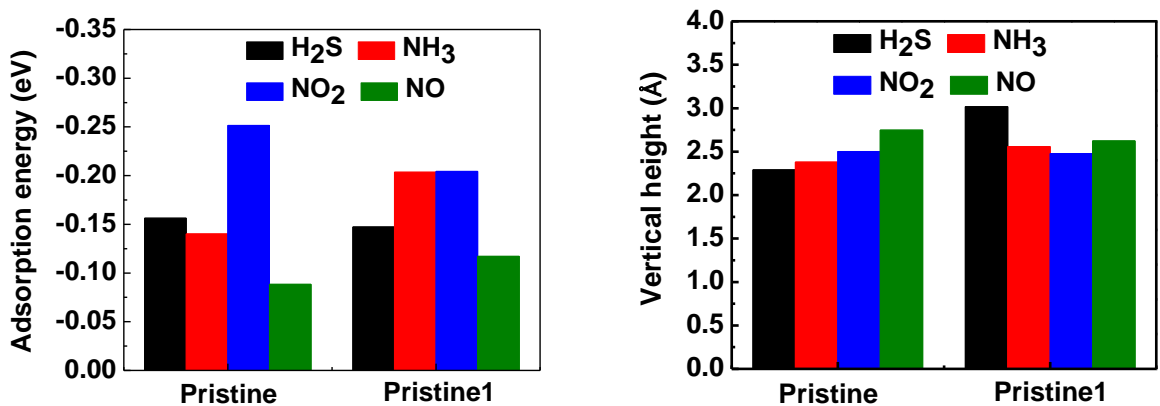

Fig.13 (a) Adsorption energy and (b) vertical height of $H_2S$, $NH_3$, $NO_2$ and NO molecule adsorbed on the surface of pristine Janus MoSSe monolayer

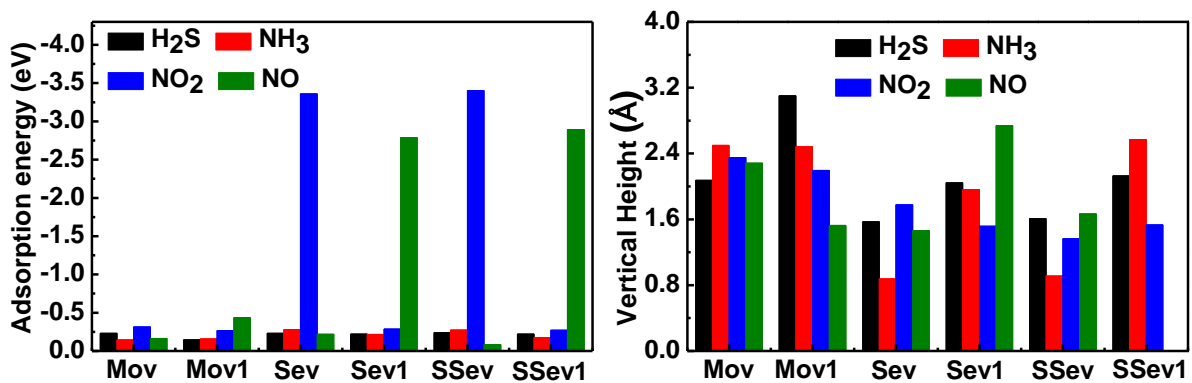

Fig.14 (a) Adsorption energy and (b) vertical height of $H_2S$, $NH_3$, $NO_2$ and NO molecule adsorbed on the surface of molybdenum vacancy, selenium vacancy and sulfur/selenium vacancy included Janus MoSSe monolayer

The adsorption energy and vertical heights are summarized in Fig 13 for pristine MoSSe monolayer in P and P1 configurations and in Fig 14 for $Mo_V$ and $Mo_V1$, $S_V$ and $S_V1$, $Se_V$ and $Se_V1$, and $S/Se_V$ and $S/Se_V1$ configurations. These studies suggest that the pristine MoSSe Janus monolayer is more sensitive to $NO_2$ molecule as compared to other considered monolayer, shown in Fig.13 (a). All considered toxic gas molecules $H_2S$, $NH_3$, $NO_2$ and NO are interacting through the weak vdW interaction, confirming the physisorption in most of the cases. The adsorption energy of molecules lies in the order of $NO_2 > NH_3 > H_2S > NO$, consistent with experimental studies by Rahul et al [58] for $MoS_2$ based sensing of these gas molecules. The adsorption energy of $NO_2$ molecule on the pristine monolayer is higher than the reported for $MoS_2$ and $WS_2$ monolayer. Further, $NO_2$ molecule takes charge from monolayer and thus, the desorption of $NO_2$ may be relatively faster. Thus, defects will play an important role to enhance the sensitivity and selectivity of transition metal dichalcogenides based gas sensor devices. The change transfer observed for $H_2S$ and $NH_3$ in the defects included MoSSe monolayers is not significant, as shown in Fig.14 (a), with respect to other gas molecules. The spin polarized band structure analysis confirms that $H_2S$ and $NH_3$ gas molecules do not show any contribution in the forbidden region and also the low charge transfer supports the physisorption for these gas molecules. $NO_2$ and NO molecules are more sensitive to defects induced Janus MoSSe monolayer. The charge transfer in case of $NO_2$ and NO gas molecules is relatively larger and in some configurations, lead to the covalent bonding with the monolayer, sharing large charge. The respective adsorption energies for $NO_2$ and NO are much larger with respect that of the other gas molecules. The large adsorption energy for $NO_2$ molecule signifies the enhanced sensitivity with respect to other gas molecules. The charge transfer mechanism defines the interaction between the adsorbent and host material that lead to the change in the resistance for practical gas sensing applications. These results suggest that the pristine and defect included MoSSe Janus monolayers will have very large sensitivity for $NO_2$ molecule as compared to the reported results for $MoS_2$ monolayer. Thus, the present study might be very useful for the experimentalist to (i) design high sensitivity MoSSe Janus layer based sensors and (ii) understand the microscopic mechanism of gas adsorption in such systems.

4. **Conclusion:**

We studied the structural properties and thermodynamic stability of pristine Janus MoSSe monolayer. Three types of vacancy defects like molybdenum, selenium and sulfur/selenium defects are considered in Janus MoSSe monolayer. The structural stability of considered defects is investigated using the formation energy and found that selenium vacancy is a most stable defect among others. Adsorption behavior of $H_2S$, $NH_3$, $NO_2$, and NO molecules are investigated and observed that $H_2S$ and $NH_3$ molecules do not show significant change in their electronic and magnetic properties. The noticed adsorption energy, vertical height, and charge transfer confirm the physisorption of $H_2S$ and $NH_3$ molecule on the pristine and defect included MoSSe monolayers. In selenium and sulfur/selenium defect included MoSSe monolayer, $NO_2$ dissociated while adsorption forming the oxygen doped MoSSe monolayer resembling like NO adsorption, with large adsorption energy. This also resulted in significant change in respective electronic and magnetic properties with the onset of about $1\mu_B$ magnetic moment after adsorption. Due to the dissociation of $NO_2$ molecule in sulfur/selenium defect included Janus monolayer can also be used as an efficient catalyst. NO molecule showed chemisorption for selenium and sulfur/selenium defect included MoSSe monolayer when NO molecule is adsorbed from the nitrogen site. Thus, very large sensitivity and selectivity for $NO_2$ and NO gas molecules in $S_V$ and $S/Se_V$ defect included MoSSe Janus monolayer may provide a novel sensing platform.